\documentclass[10pt,aps,twocolumn,prd,notitlepage,amssymb,amsmath,floatfix,nofootinbib,superscriptaddress]{revtex4-1}

\usepackage{epsfig}
\usepackage{bm}
\usepackage{amssymb}
\usepackage{amsmath}
\usepackage{physics}
\usepackage{color}
\usepackage{framed}
\usepackage{xcolor}
\usepackage{subfigure}
\usepackage[colorlinks,linkcolor=blue,anchorcolor=black,citecolor=blue]{hyperref}
\usepackage{multirow}

\allowdisplaybreaks[4]

\mathchardef\mhyphen="2D

\makeatother

\begin{document}

\title{A systematic study of global spin polarizations and correlations of hadrons with different spins in relativistic heavy ion collisions}

\author{Ji-peng Lv}
\email{jipenglv@mail.sdu.edu.cn}
\affiliation{Institute of Frontier and Interdisciplinary Science, Key Laboratory of Particle Physics and Particle Irradiation (MOE), Shandong University, Qingdao, Shandong 266237, China}

\author{Zi-han Yu}
\email{zihan.yu@mail.sdu.edu.cn}
\affiliation{Institute of Frontier and Interdisciplinary Science,  Key Laboratory of Particle Physics and Particle Irradiation (MOE), Shandong University, Qingdao, Shandong 266237, China}

\author{Xiao-wen Li}
\email{xiaowen.li@sdu.edu.cn}
\affiliation{Institute of Frontier and Interdisciplinary Science, Key Laboratory of Particle Physics and Particle Irradiation (MOE), Shandong University, Qingdao, Shandong 266237, China}

\author{Zuo-tang Liang}
\email{liang@sdu.edu.cn}
\affiliation{Institute of Frontier and Interdisciplinary Science, Key Laboratory of Particle Physics and Particle Irradiation (MOE), Shandong University, Qingdao, Shandong 266237, China}

\begin{abstract}
Based on the formalism presented in~\cite{Lv:2024uev}, we make a systematic study of spin polarizations of hadrons 
with different spins including vector mesons, spin-1/2 and spin-3/2 hyperons
and spin correlations of hadron-hadron such as hyperon-hyperon, hyperon-vector meson and vector meson-vector meson in relativistic heavy ion collisions.  
We present the results obtained and discuss the physical consequences in special cases. 
These results can be used for further numerical studies and for references of future experiments as well.
\end{abstract}

\maketitle

\section{Introduction}

The global polarization effect~\cite{Liang:2004ph,Liang:2004xn,Gao:2007bc} in non-central relativistic heavy ion collisions 
due to spin–orbit coupling in strong interactions provides valuable insights into the properties of the quark-gluon plasma (QGP). 
The effect has been observed not only for 
hyperon polarizations~\cite{STAR:2017ckg,STAR:2018gyt,STAR:2020xbm,STAR:2021beb,ALICE:2019onw,HADES:2022enx,STAR:2023nvo}, 
but also for vector meson spin alignments~\cite{STAR:2022fan,STAR:2008lcm,ALICE:2019aid,ALICE:2022byg,ALICE:2022dyy} 
and have attracted much attention 
~\cite{Betz:2007kg,Becattini:2007sr,Karpenko:2016jyx,Pang:2016igs,Li:2017slc,Xie:2017upb,Sun:2017xhx,Baznat:2017jfj,Shi:2017wpk,Xia:2018tes,Wei:2018zfb,Fu:2020oxj,Ryu:2021lnx,Fu:2021pok,Deng:2021miw,Becattini:2021iol,Wu:2022mkr} 
(see also reviews~\cite{Gao:2020lxh,Huang:2020dtn,Becattini:2021lfq,Florkowski:2018fap,Becattini:2020ngo,Gao:2020vbh,Becattini:2022zvf,Jian-Hua:2023cna,Liang:2019clf,Liang:2007ma,Wang:2017jpl,Huang:2020xyr,Liang:2022ekv,Wang:2024wul,Huang:2024ffg,Becattini:2024uha}). 
The data show in particular that the produced quark matter system, the QGP,  is not only globally polarized but also possesses  strong spin correlations~\cite{Lv:2024uev,Zhang:2024hyq}.
This has led to rapid new developments in theoretical studies~\cite{Yang:2017sdk,Sheng:2019kmk,Sheng:2020ghv,Sheng:2022ffb,Sheng:2022wsy,Chen:2024afy,Sheng:2026hsc}, 
marking a new phase in spin physics in heavy ion collisions~\cite{Wang:2023fvy,Chen:2023hnb,talks,Xin-Li:2023gwh,Li-Juan:2023bws,Xia:2020tyd,Wei:2023pdf,Kumar:2023ghs,DeMoura:2023jzz,Fu:2023qht,Sheng:2023urn,Fang:2023bbw,Dong:2023cng,Kumar:2023ojl,Augustin:1978wf,Chen:2016iey,Xu:2024kdh,Wang:2024tsr,Hua:2024bwn,Oliva:2026wbo}.
The study has been extended even to spin correlations in other high energy reactions such as $e^+e^-$, $ep$, and $pp$ collisions~\cite{Zhang:2023ugf,Li:2023qgj,Huang:2024awn},  
as well as the QCD critical end point~\cite{Chen:2024hki} and quantum entanglement effects~\cite{Wu:2024mtj,Wu:2024asu} in heavy ion collisions.

A systematic framework to describe spin polarizations and correlations in quark matter systems and their relationships to observables at the hadron-level has recently been developed in~\cite{Lv:2024uev}. 
%
In view of the growing interests in spin correlations, we make a comprehensive study of spin polarizations of hadrons with different spins including vector mesons, spin-1/2 and spin-3/2 hyperons
and spin correlations between two hadrons including hyperon-hyperon, hyperon-vector meson and vector meson-vector meson in relativistic heavy ion collisions using the formalism given in~\cite{Lv:2024uev}. 
We unify the definitions and present the systematic results in this paper. 
The results can be used for numerical analysis and as references for experiments. 

The rest of the paper is organized as follows. 
In Sec.~\ref{sec:formulae}, we briefly review the formalism developed in~\cite{Lv:2024uev} and present the complete results for hadrons in Sec.~\ref{sec:results}. 
We make a short summary and discussion in Sec.~\ref{sec:summary}. 

\section {The calculation formulae} \label{sec:formulae}

As  in~\cite{Lv:2024uev}, we consider a quark matter system consisting of quarks and antiquarks produced in relativistic heavy ion collisions 
and assume that they hadronize via the quark combination mechanism. 
We summarize the key formulae to describe quark spin polarizations and correlations of the system 
and to calculate the relationships between them and those for the produced hadrons. 
We extend the formulae in particular to spin correlations of hadron-hadron such as hyperon-vector meson and vector meson-vector meson in a unified way.    
 
 \subsection{Quark spin polarizations and correlations}
 \label{sec:quark}
For the quark system,  the spin density matrix for single quark (anti-quark) is given by
\begin{align}
    \hat{\rho}^{(q)}= \frac{1}{2}(1+P_{qi} \hat{\sigma}_{i}),
    \label{eq:Pqdef}
\end{align}
where $P_{qi}=\langle \hat{\sigma}_i \rangle=\text{Tr}\hat{\rho}^{(q)} \hat{\sigma}_i$ with $i=1-3$ (representing $x,y,z$) is the $i$-th component of the quark polarization, 
$\hat{\sigma}_i$ is the Pauli matrix and $\hat{\rho}^{(q)}$ is normalized as  $\text{Tr} \hat{\rho}^{(q)}=1$.  
Here as well as in the following of this paper, we use the convention that repeated indices are summed.

For two particle system (12), $\hat{\rho}^{(12)}$ is expanded as
\begin{align}
    \hat{\rho}^{(12)}=\hat{\rho}^{(1)} \otimes  \hat{\rho}^{(2)} +\frac{1}{2^2} c^{(12)}_{ij} \hat{\sigma}_{1i} \otimes \hat{\sigma}_{2j},
    \label{eq:qcijdef}
\end{align}
where $c^{(12)}_{ij}$ is the spin correlation between the two particles 1 and 2. 
We note in particular that $c^{(12)}_{ij}=0$ if there is no spin correlation between them.

If other degrees of freedom, denoted by $\alpha$, of the particle are considered, we generalize the spin density matrices to include the 
$\alpha$-dependence, i.e., 
\begin{align}
    \hat{\rho}^{(1)}(\alpha_1)=&\frac{1}{2}[1+P_{1i}(\alpha_1) \hat{\sigma}_{1i} ],
    \label{eq:Palpha}
    \\
    \hat{\rho}^{(12)}(\alpha_1,\alpha_2)=& \hat{\rho}^{(1)}(\alpha_1)
    \otimes \hat{\rho}^{(2)}(\alpha_2)
    \nonumber \\
    &+\frac{1}{2^2} c^{(12)}_{ij}(\alpha_1,\alpha_2)\hat{\sigma}_{1i}
    \otimes \hat{\sigma}_{2j}.
    \label{eq:cijalpha}
\end{align}

If the system (12) is at the state $\ket{\alpha_{12}}$, we often consider the effective spin density matrix given by
\begin{align}
    \hat{\bar{\rho}}^{(12)}(\alpha_{12})&=\langle \alpha_{12}|\hat{\rho}^{(12)}|\alpha_{12} \rangle\equiv\langle\hat{\rho}^{(12)}\rangle \nonumber\\
    &=\sum_{\alpha_1,\alpha_2} \hat{\rho}^{(12)} (\alpha_1,\alpha_2) |\langle \alpha_1,\alpha_2|\alpha_{12} \rangle|^2,
    \label{eq:cijeffective}
\end{align}
that is the average over $(\alpha_1,\alpha_2)$ in the state $|\alpha_{12} \rangle$. 
Here, it is interesting to note that the same decomposition scheme as given by Eqs.~(\ref{eq:Pqdef}) and (\ref{eq:qcijdef}), i.e., 
\begin{align}
    \hat{\bar{\rho}}^{(1)}=& \frac{1}{2}(1+\bar{P}_{1i} \hat{\sigma}_{1i}),
    \label{eq:Pbar}
    \\
    \hat{\bar{\rho}}^{(12)}=& \hat{\bar{\rho}}^{(1)} \otimes \hat{\bar{\rho}}^{(2)}+\frac{1}{2^2} \bar{c}^{(12)}_{ij}
    \hat{\sigma}_{1i} \otimes \hat{\sigma}_{2j},
    \label{eq:cijbar}
\end{align}
leads to the effective polarization $\bar{P}_{1i}$ and correlation $\bar{c}^{(12)}_{ij}$ as 
$\bar{P}_{1i}=\langle P_{1i} \rangle$, but $\bar{c}^{(12)}_{ij}\not=\langle c^{(12)}_{ij} \rangle$, instead 
\begin{align}
  \bar{c}^{(12)}_{ij}=& \langle c^{(12)}_{ij} \rangle+\bar{c}^{(12;0)}_{ij},  
  \label{eq:cijbar2}
  \\
  \bar{c}^{(12;0)}_{ij}=& \langle P_{1i} P_{2j} \rangle- \bar{P}_{1i} \bar{P}_{2j,}
    \label{eq:cij0}
\end{align}
where  
$c^{(12)}_{ij}$ and $\bar{c}^{(12;0)}_{ij}$
are referred to as the genuine and induced quark spin correlations respectively~\cite{Lv:2024uev}.

It is straightforward to apply the formalism to systems of more than two particles~\cite{Lv:2024uev} and we omit them here.

\subsection{Hadron spin polarizations} \label{sec:hpol}

The hadron spin polarizations are determined by calculating the corresponding spin density matrices. 

For spin-1/2 hadrons such as hyperons $H$ in the $J^P=(1/2)^+$ baryon octet, 
the spin density matrix is decomposed in exactly the same way as that for quarks given by Eq.~(\ref{eq:Pqdef}), i.e., 
\begin{align}
    \hat{\rho}^{H}= \frac{1}{2}(1+P_{Hi} \hat{\sigma}_{i}),
    \label{eq:Hdef}
\end{align}
where $P_{Hi}=\langle\hat\sigma_i\rangle=\Tr\hat\rho^H\hat\sigma_i$ is the polarization of $H$ and, in terms of elements of $\hat\rho^H$, is given by
\begin{align}
&P_{Hx}=2\Re\rho_{+-}^H=2\Re\rho_{-+}^H, \\  
&P_{Hy}=-2\Im\rho_{+-}^H=2\Im\rho_{-+}^H, \\  
&P_{Hz}=\rho_{++}^H-\rho_{--}^H. 
\end{align}

For spin-1 hadrons such as vector mesons $V$, the spin state is described by a $3\times3$ spin density matrix.
We often simply use the corresponding matrix elements to describe the spin properties such as $\rho^V_{00}$ to describe the spin alignment~\cite{Liang:2004xn}. 
Consistent with that for spin-1/2 hadrons, we also decompose it in terms of a set of $3\times 3$ hermitian matrices. 
A commonly used way is~\cite{Bacchetta:2000jk,Chen:2015ora,Chen:2016moq},
\begin{align}
\hat{\rho}^V = \frac{1}{3} (\mathbf{1} + \frac{3}{2}S^i_V \Sigma_i + 3 T_V^{ij} \Sigma_{ij}), \label{eq:spin1rho}
\end{align}
where $\Sigma_i$ with a subscript $i=x,y,z$ is the spin operator of a spin-$1$ particle, 
and $\Sigma_{ij}= \frac{1}{2} (\Sigma_i\Sigma_j + \Sigma_j \Sigma_i) - \frac{2}{3} \mathbf{1} \delta_{ij}$.  
In this scheme, the polarization of vector $V$ is described by a polarization vector $S_V=(0,S_{V}^i)$ 
and a tensor polarization part $T^{ij}_V$ that has five independent components including    
a Lorentz scalar $S_{LL}$, a vector $S_{LT}^\mu = (0, S_{LT}^x, S_{LT}^y,0)$, 
and a tensor $S_{TT}^{\mu\nu}$ that has two nonzero independent components 
$S_{TT}^{xx} = -S_{TT}^{yy}$ and $S_{TT}^{xy} = S_{TT}^{yx}$. 
We write it in the same form as Eq.~(\ref{eq:Hdef}), i.e., 
\begin{align}
    \hat{\rho}^{V}= \frac{1}{3}(1+\gamma_V^{(l)}P_{V}^{(l)} \hat{\Sigma}_V^{(l)}).
    \label{eq:Vdef}
\end{align}
Here $P_{V}^{(l)}$ $(l=1-8)$ denote  the eight independent components describing the polarization of $V$,  
and $\hat\Sigma_{V}^{(l)}$ are the 8 corresponding independent orthogonal traceless $3\times 3$ hermitian matrices.  
More precisely, they are given by
\begin{align}
&P_{V}^{(l)}=(S_V^i,S_{LL},S_{LT}^{a},S_{TT}^{x a}), \\ 
&\hat\Sigma_{V}^{(l)}=(\Sigma^i,\hat\Sigma_{LL},\hat\Sigma_{LT}^{a},\hat\Sigma_{TT}^{x a}), 
\end{align}
where $l=1-8$, $i=1-3$, and $a=1-2$; 
$\Tr \hat\Sigma_{V}^{(l)}=0$ and  $\Tr \hat\Sigma_{V}^{(l)} \hat\Sigma_{V}^{(j)}=\delta_{lj}\Tr \hat\Sigma_{V}^{(l)2}$; 
$\hat\Sigma_{LL}=3\Sigma_{zz}/2$, $\hat\Sigma_{LT}^{a}=2\Sigma_{z a}$, $\hat\Sigma_{TT}^{xx}=\Sigma^2_x-\Sigma_y^2, \hat\Sigma_{TT}^{xy}=2\Sigma_{xy}$; 
$\Tr\hat\Sigma_V^{(l)2}=2$ for $l\not= LL$ and $\Tr\hat\Sigma_{LL}^2=3/2$. 
The factor $\gamma_V^{(l)}=3/{\Tr\hat\Sigma_{V}^{(l)2}}$ is introduced to guarantee that $P_{V}^{(l)}$ takes the form as
\begin{align}
&P_{V}^{(l)}=\langle \hat\Sigma_{V}^{(l)}\rangle =\Tr [\hat\rho^V\hat\Sigma_{V}^{(l)}]. 
\end{align}
We see that $\gamma_V^{(l)}=3/2$ for $l\not= LL$ and $\gamma_{LL}=2$. 

In terms of elements of $\hat\rho^V$,  the five tensor polarization components of $P_{V}^{(l)}$ are given by, 
\begin{align} 
S_{LL}=&(1-3\rho^V_{00})/2,\\
S_{LT}^x=&\sqrt{2}\Re (\rho^V_{10}-\rho^V_{0-1}),\\
S_{LT}^y=&-\sqrt{2}\Im (\rho^V_{10}-\rho^V_{0-1}),\\
S_{TT}^{xx}=&2\Re\rho^V_{1-1}=2\Re\rho^V_{-11},\\
S_{TT}^{xy}=&-2\Im\rho^V_{1-1}=2\Im\rho^V_{-11}.
\end{align}

We also note that, similar to hyperon polarizations, polarizations of vector mesons are mainly measured via angular distributions of decay products of $V\to MM$ (two spinless mesons). 
In this way, we can only measure tensor polarizations of vector mesons. 
In particular if we study only polar angle $\theta$ distributions, we measure only $S_{LL}$. 
See appendix~\ref{appendix:decay} for details.  
Hence, if we want to study whether the tensor polarization is isotropic, we need either to study both polar and azimuthal angular distributions 
or choose different quantization axises of spin and study the corresponding $S_{LL}$ respectively.

For spin-3/2 baryons, the spin density matrix $\hat\rho^{H_3}$ is $4\times 4$. 
We have 15 independent components of spin polarization and corresponding traceless hermitian matrices $\hat\Sigma^{(l)}_{H_3}$. 
The polarization is decomposed into a vector, a rank-2 and a rank-3 tensor part, i.e., 
\begin{align}
&P_{H_3}^{(l)}=(S^i,S_{LL},S_{LT}^{a},S_{TT}^{x a},S_{LLL},S_{LLT}^{a},S_{LTT}^{x a},S_{TTT}^{xx a}), \nonumber\\ 
&\hat\Sigma_{H_3}^{(l)}=(\hat\Sigma^i,\hat\Sigma_{LL},\hat\Sigma_{LT}^{a},\hat\Sigma_{TT}^{x a},\hat\Sigma_{LLL},\hat\Sigma_{LLT}^{a},\hat\Sigma_{LTT}^{x a},\hat\Sigma_{TTT}^{xx a}), \nonumber
\end{align}
and $\hat\rho^{H_3}$ is expanded in terms of $\hat\Sigma^{(l)}_{H_3}$ in the same form as given by Eqs.~(\ref{eq:Hdef}) and (\ref{eq:Vdef}). 
More details can be found e.g. in \cite{Zhang:2024hyq,Kim:1992az,Zhao:2022lbw} and we do not repeat here. 

To summarize, for hadron with different spins, we unify the decomposition of the spin density matrix by
\begin{align}
    \hat{\rho}^{h}= \frac{1}{2j_h+1}(1+\gamma_h^{(l)}P_{h}^{(l)} \hat{\Sigma}_h^{(l)}),
    \label{eq:hdef}
\end{align}
where $l=1-n_h$ and $n_h=(2j_h+1)^2-1$ is the number of independent real components of $\hat{\rho}^{h}$ for hadron $h$ with spin $j_h$;
$\hat{\Sigma}_h^{(l)}$ is a complete set of traceless hermitian matrices in the corresponding spin space;  
$\gamma_h^{(l)}=(2j_h+1)/\Tr\hat{\Sigma}_h^{(l)2}$ and different components of hadron spin polarizations are given by 
\begin{align}
    P_h^{(l)}=\langle \hat{\Sigma}_h^{(l)}\rangle=\Tr\hat\rho^h\hat\Sigma_h^{(l)}. 
    \label{eq:Phdef}
\end{align}

In the quark combination mechanism, the spin density matrix $\hat\rho^h$ of the hadron is determined by that of quarks (antiquarks) in the quark matter system 
and the transition operator $\hat{\cal M}$, i.e.,  
\begin{align}
\hat\rho^{h}=\hat{\cal M}\hat \rho^{([q_n])}\hat{\cal M}^\dag, \label{eq:htransition}
\end{align}
where $[q_n]$ denotes $q_1\bar{q}_2$ if $h$ is a meson and $q_1q_2q_3$ if $h$ is a baryon. 

Here, it is important to note that spin density matrix elements of the produced hadron $h$  are independent of the transition operator $\hat{\cal M}$ due to space rotation invariance~\cite{Lv:2024uev} 
in the case that only spin degree of freedom is considered. 
More precisely, though $\hat\rho^{h}$ depends on $\hat{\cal M}$ as given by Eq.~(\ref{eq:htransition}), 
the matrix element is given by
\begin{align}
   \rho^{h}_{m_h m_h^{\prime}} 
  =&N \sum_{m_n m_n^{\prime}} \langle j_hm_h|m_n \rangle \langle m_n^{\prime} |j_hm_h^{\prime} \rangle \rho^{([q_n])}_{m_n m_n^{\prime}},  
    \label{eq:rhoh}
\end{align}
where $|j_hm_h\rangle$ is the spin state of $h$, $\ket{m_n }\equiv\ket{j_1 m_1j_2 m_2}$ denotes the spin state of $q_1\bar q_2$ if $h$ is a meson,  
it is $\ket{j_1 m_1j_2 m_2j_3m_3}$ of $q_1q_2q_3$ if $h$ is a baryon; and $\langle j_hm|m_n \rangle$ is just the Clebsch-Gordan coefficient.
The contribution of $\hat{\cal M}$ depends only on the total angular momentum $j_h$ thus is absorbed into the normalization constant. 
It does not contribute to $\rho^h_{m_hm_h^{\prime}}$ but contribute to the relative production weights of hadrons with different spins.

If $\alpha$-dependence is taken into account, the conclusion remains if the hadron wave function is factorized. 
In this case we only need to replace the spin matrix elements for the quark (antiquark) system by the effective ones, i.e.  
\begin{align}
\rho^{h}_{m_h m_h^{\prime}} (\alpha_h)
=&N\sum_{m_n m_n^{\prime}} \bar\rho^{([q_n])}_{m_n m_n^{\prime}}(\alpha_h)\nonumber\\
 &\times \langle j_hm_h|m_n \rangle  \langle m_n^{\prime} |j_hm_h^{\prime} \rangle.  
\label{eq:halpha}
\end{align}
Here, we emphasize that $\bar\rho^{([q_n])}_{m_n m_n^{\prime}}$ in Eq.~(\ref{eq:halpha}) is given by 
\begin{align}
\bar\rho^{([q_n])}_{m_n m_n^{\prime}}(\alpha_h)&=\langle\alpha_h|\rho^{([q_n])}_{m_n m_n^{\prime}}(\alpha_{q_n})|\alpha_h\rangle,  
\label{eq:barrho}
\end{align}
where the average is carried out inside the hadron $h$ and is in general different for different hadrons. 
We note also that the corresponding spin correlations involved are called ``local" in~\cite{Lv:2024uev} because the average is limited to be inside a single hadron. 

\subsection{Hadron-hadron spin correlations} \label{sec:hh}

To unify, we define the spin correlation between two hadrons $h_1h_2$ via the decomposition of the spin density matrix $\hat\rho^{h_1h_2}$ 
in the same way as that given by Eq.~(\ref{eq:qcijdef}) for quarks and/or antiquarks. 
More precisely, for $h_1$ and $h_2$ with difference spins, we decompose,
\begin{align}
    \hat{\rho}^{h_1h_2}= & \hat{\rho}^{h_1} \otimes  \hat{\rho}^{h_2}  \nonumber\\ 
    &+\frac{\gamma_{h_1}^{(i)}\gamma_{h_2}^{(j)}  }{(2j_{h_1}+1)(2j_{h_2}+1)} c^{h_1h_2}_{ij}
     \hat{\Sigma}_{h_1}^{(i)} \otimes \hat{\Sigma}_{h_2}^{(j)},  \label{eq:hcijdef}
\end{align}
where $\hat{\rho}^{h_i} $ is the spin density matrix of $h_i$ given by Eq.~(\ref{eq:hdef}), $c^{h_1h_2}_{ij}$ is the spin correlation between the two hadrons $h_1$ and $h_2$ and is given by, 
\begin{align}
c^{h_1h_2}_{ij}=\langle \hat{\Sigma}_{h_1}^{(i)} \hat{\Sigma}_{h_2}^{(j)}\rangle-P_{h_1}^{(i)}P_{h_2}^{(j)}. \label{eq:cijcal}
\end{align}
We note in particular that $c^{h_1h_2}_{ij}$ defined this way satisfies $c^{h_1h_2}_{ij}=0$ if there is no spin correlation between $h_1$ and $h_2$.
This differs from the traditional definition by the addend $P_{h_1}^{(i)}P_{h_2}^{(j)}$. 
We use this unified definition in the following of this paper and suggest to use it because the physical significance is more obvious and it is easier to extend to hadrons of different spins. 

It is straightforward to find out the expression of the spin correlation $c^{h_1h_2}_{ij}$ in terms of the spin density matrix element $\rho^{h_1h_2}_{m_{h_1}m_{h_2},m_{h_1}^\prime m_{h_2}^\prime}$.  
We consider three representative systems: $h_1 h_2 = H_1 H_2$ (two hyperons), $H V$ (a hyperon and a vector meson) and $V_1 V_2$ (two vector mesons), which in general contain 9, 24 and 56 independent correlation components respectively.
However, similar to vector meson polarizations, if we measure them via polar angular distributions of decay products, only $LL$ related components can be measured 
when vector mesons are involved. 
Therefore, in the following of this paper we concentrate on these $LL$ correlations and present illustrative examples.

The most simple case is that $h_1h_2=H_1H_2$.
We denote $t_{ij}^{h_1h_2} =c_{ij}^{h_1h_2}+P_{h_1}^{(i)}P_{h_2}^{(j)}$, which combines the genuine spin correlation with the product of single-particle polarizations, and we have, e.g., 
\begin{align}
t_{zz}^{H_1H_2} = & \rho_{++,++}^{H_1H_2}+\rho_{--,--}^{H_1H_2}-\rho_{+-,+-}^{H_1H_2}-\rho_{-+,-+}^{H_1H_2}, \label{eq:tzzH1H2} \\
t_{xx}^{H_1H_2} = & 2\Re[\rho_{-+,+-}^{H_1H_2}+\rho_{--,++}^{H_1H_2}], \label{eq:txxH1H2} \\
t_{zx}^{H_1H_2} = & 2\Re[\rho_{+-,++}^{H_1H_2}-\rho_{--,-+}^{H_1H_2}], \label{eq:tzxH1H2} \\
t_{xz}^{H_1H_2} = &2\Re[\rho_{-+,++}^{H_1H_2}-\rho_{--,+-}^{H_1H_2}] . \label{eq:txzH1H2} 
\end{align}

For $h_1h_2=HV$ or $V_1V_2$, we have, e.g., 
\begin{align}
t_{z,LL}^{HV} =\frac{1}{2} & \bigl( \rho^{HV}_{+1,+1}+\rho^{HV}_{+-1,+-1} -2\rho^{HV}_{+0,+0} \nonumber\\ 
 &-\rho^{HV}_{-1,-1}-\rho^{HV}_{--1,--1} +2\rho^{HV}_{-0,-0}\bigr), \label{eq:tzLLHV} \\
t_{LL,LL}^{V_1V_2} = \frac{1}{4} & \bigl[ \rho^{V_1V_2}_{11,11}-2\rho^{V_1V_2}_{10,10}+\rho^{V_1V_2}_{1-1,1-1}-2\rho^{V_1V_2}_{01,01}
\nonumber
\\
&+4\rho^{V_1V_2}_{00,00}-2\rho^{V_1V_2}_{0-1,0-1}+\rho^{V_1V_2}_{-11,-11}
\nonumber
\\
&-2\rho^{V_1V_2}_{-10,-10}+\rho^{V_1V_2}_{-1-1,-1-1}
\bigr]. \label{eq:tLLLLVV} 
\end{align}

We also note that these hadron-hadron spin correlations can be measured via angular distributions of decay products in similar way as those for hyperon polarizations. 
We summarize the key formulae relating these spin correlation components to the measurable decay angular distributions in Appendix~\ref{appendix:decay},
where we see explicitly that similar to vector meson polarizations, only $LL$ related spin correlations between vector mesons and others can be measured via polar angular $\theta$ distributions of vector meson to two spinless mesons.

In the quark combination mechanism, the spin density matrix element of $h_1h_2$ is given by
\begin{align}
&\rho^{h_1h_2}_{m_{h_i} m_{h_i}^{\prime}} 
=N\sum_{m_{h_i} m_{h_i}^{\prime}}  \bar\rho^{([q_n])}_{m_n m_n^{\prime}} 
\langle m_{h_i}|m_n \rangle  \langle m_n^{\prime} |m_{h_i}^{\prime} \rangle .  
\label{eq:h1h2alpha}
\end{align}
Here the subscript $i=1,2$ of ${h_i}$ stands for the two hadrons and $|m_{h_i}\rangle$ represents $| j_{h_1}m_{h_1} j_{h_2}m_{h_2}\rangle$;  
$\langle m_n |m_{h_i} \rangle$ is the product of two Clebsch–Gordan coefficients $\langle m_{h_1}| m_{n_1} \rangle$ and $\langle m_{h_2}| m_{n_2} \rangle$,  
and the contribution of the transition operator $\hat{\mathcal{M}}$ is again absorbed into the normalization constant. 
This is valid if $\hat{\mathcal{M}}=\hat{\mathcal{M}}_{h_1} \hat{\mathcal{M}}_{h_2}$, i.e., factored to the product of that for $h_1$ and $h_2$ separately. 

The effective spin density matrix for the quarks (antiquarks) $[q_n]$ is given by
\begin{align}
\hat{\bar\rho}^{([q_n])}&(\alpha_{h_1},\alpha_{h_2})
=\langle\alpha_{h_1}\alpha_{h_2}|\hat\rho^{([q_n])}(\alpha_{q_n})|\alpha_{h_1}\alpha_{h_2}\rangle\nonumber\\
=&\sum_{\alpha_{q_a}\alpha_{q_b}}  |\langle\alpha_{q_a}\alpha_{q_b}|\alpha_{h_1}\alpha_{h_2}\rangle|^2 \hat\rho^{([q_n])}(\alpha_{q_a},\alpha_{q_b}),  
\label{eq:rhoqneffective}
\end{align}
where $q_a$ and $q_b$ denote quarks (antiquarks) that combine into $h_1$ and $h_2$ respectively. 
If we do not consider the overlap of the wave function of $h_1$ with that of $h_2$ in the $\alpha$-space, we have   
$\langle\alpha_{q_a}\alpha_{q_b}|\alpha_{h_1}\alpha_{h_2}\rangle=\langle\alpha_{q_a}|\alpha_{h_1}\rangle\langle\alpha_{q_b}|\alpha_{h_2}\rangle$, 
and we obtain
\begin{align}
\hat{\bar\rho}^{([q_n])} (\alpha_{h_1},\alpha_{h_2})
=\sum_{\alpha_{q_a}\alpha_{q_b}} &  |\langle\alpha_{q_a}|\alpha_{h_1}\rangle|^2|\langle\alpha_{q_b}|\alpha_{h_2}\rangle|^2 \nonumber\\ 
\times&\hat\rho^{([q_n])}(\alpha_{q_a},\alpha_{q_b}).  
\label{eq:rhoqneffective2}
\end{align}
Correspondingly the average quark spin correlation 
\begin{align}
&\bar c_{ij}^{(q_aq_b)} (\alpha_{h_1},\alpha_{h_2})=\langle c_{ij}^{(q_aq_b)}(\alpha_{q_a},\alpha_{q_b})\rangle_{h_1h_2} \nonumber \\
&=\sum_{\alpha_{q_a}\alpha_{q_b}} |\langle\alpha_{q_a}|\alpha_{h_1}\rangle|^2|\langle\alpha_{q_b}|\alpha_{h_2}\rangle|^2  
 c_{ij}^{(q_aq_b)}(\alpha_{q_a},\alpha_{q_b}),  
\label{eq:cijeffectiveh1h2}
\end{align}
has only long range contribution because $q_a$ and $q_b$ belong to $h_1$ and $h_2$ respectively. 
Furthermore, there is no contribution from induced correlations in this case because  

\begin{align}
\langle P_{q_ai}P_{q_bj} \rangle_{h_1h_2}
=&\sum_{\alpha_{q_a}\alpha_{q_b}}  |\langle\alpha_{q_a}|\alpha_{h_1}\rangle|^2|\langle\alpha_{q_b}|\alpha_{h_2}\rangle|^2 \nonumber\\ 
&~~~~~~~\times P_{q_ai}(\alpha_{q_a})P_{q_bj}(\alpha_{q_b})\nonumber\\
=&\bar P_{q_ai}(\alpha_{h_1}) \bar P_{q_bj}(\alpha_{h_2}).
\label{eq:P1P2effectiveh1h2}
\end{align}   

We emphasize that the above mentioned results are valid for two identical hadrons if we neglect the overlap in $\alpha$-space for 
the wave functions of these two hadrons. 
We consider only this case in the present paper. 

We also emphasize that the spin correlations calculated from Eq.~(\ref{eq:h1h2alpha}) are for hadrons $h_1$ and $h_2$ 
at given $\alpha_{h_1}$ and $\alpha_{h_2}$ respectively thus are functions of $(\alpha_{h_1},\alpha_{h_2})$. 
In practice, we need to average over $(\alpha_{h_1},\alpha_{h_2})$ for $h_1$ and $h_2$ in the system produced in high energy collisions. 
The average should be made for the spin density matrix $\hat\rho^{h_1h_2}$ thus induced spin correlations may contribute also at this level.

\section{The calculation results} \label{sec:results}

Using the formulae given in Sec.~\ref{sec:formulae}, we calculate spin polarizations and correlations of hadrons with different spins. 
We obtain results in terms of those of quarks and/or anti-quarks in the quark system. 
Results for spin polarizations of hadrons and those for $\Lambda\bar\Lambda$ spin correlations in the case that only two quark spin correlations are considered 
have been given in~\cite{Lv:2024uev} and \cite{Zhang:2024hyq}. 
In the following of this section, we present complete results for spin correlations of hadrons with different spins such as spin-1/2, 1 and 3/2 respectively. 
For comparison, we include also some of results for spin polarizations of hadrons.  
To be explicit, we present results in the case that $\alpha$-dependence is considered 
so that the quark spin polarization and  correlation are averaged results given by Eqs.~(\ref{eq:barrho}) and (\ref{eq:rhoqneffective}) respectively 
and denoted by a bar over the symbols. 
We first present results in the general case and discuss some special cases at the end of this section. 

\subsection{Hadron spin polarizations} \label{sec:hres}

{\it Spin-1/2 hyperons:}
The results for $J^P=(1/2)^+$ hyperons are summarized as follows.  
\begin{align}
 &P_{\Lambda j}(\alpha_{\Lambda})=\bar{P}_{sj}+\frac{\delta\bar\rho_{\Lambda j}}{\bar{C}_{\Lambda}},       \label{eq:PLambda}  \\  
 &P_{\Sigma^0 j}(\alpha_{\Sigma^0})=\frac{1}{3}(2\bar{P}_{uj}+2\bar{P}_{dj}-\bar{P}_{sj})+\frac{\delta \bar{\rho}_{\Sigma^0j}}{\bar{C}_{\Sigma^0}}, \label{eq:PSigma0} \\
 &P_{H_{112}j}(\alpha_{H_{112}})=\frac{1}{3}(4  \bar{P}_{q_1j}-  \bar{P}_{q_2j})+\frac{  \delta \bar{\rho}_{H_{112}j}}{  \bar{C}_{H_{112}}}, \label{eq:PH112}
\end{align}
where $j = x, y, z$; $H_{112}$ denotes hyperons with quark flavor content $q_1q_1q_2$ such as $\Sigma^\pm$ and $\Xi$. 
The numerators $\delta \bar{\rho}_{Hj}$'s are given by
\begin{align}
  \delta \bar{\rho}_{\Lambda j}=&\bar{P}_{sj} \bar{t}^{(ud)}_{ii}-\bar{t}^{(uds)}_{iij},\\
  \delta \bar{\rho}_{\Sigma^0j}=&\bar{t}^{(uds)}_{iij}-2\bar{t}^{(uds)}_{jii}-2\bar{t}^{(uds)}_{iji}  \nonumber  \\
   &-\frac{1}{3}(\bar{t}^{(ud)}_{ii}-2\bar{t}^{(us)}_{ii}-2\bar{t}^{(ds)}_{ii})
    \nonumber \\
    &\quad \times (2\bar{P}_{uj}+2 \bar{P}_{dj}-\bar{P}_{sj}),    \\
    \delta \bar{\rho}_{H_{112}j}=&\bar{t}^{(q_1 q_1 q_2)}_{iij}-4\bar{t}^{(q_1 q_1 q_2)}_{jii} \nonumber  \\
   &-\frac{1}{3}(\bar{t}^{(q_1 q_1)}_{ii}-4\bar{t}^{(q_1 q_2)}_{ii})(4\bar{P}_{q_1j}- \bar{P}_{q_2 j}), 
\end{align}
where $\bar{t}^{(q_1 q_2)}_{ij}$ and $\bar{t}^{(q_1 q_2 q_3)}_{ijk}$ are contributions from two and three quark spin correlations given by
\begin{align}
    \bar{t}^{(q_1 q_2)}_{ij} =&  \bar{c}^{(q_1 q_2)}_{ij}+\bar{P}_{q_1 i} \bar{P}_{q_2 j},  \label{eq:tq1q2}     \\
    \bar{t}^{(q_1 q_2 q_3)}_{ijk} =& \bar{c}^{(q_1 q_2 q_3)}_{ijk}+
    \bar{c}^{(q_1 q_2)}_{ij} \bar{P}_{q_3 k}+\bar{c}^{(q_2 q_3)}_{jk} \bar{P}_{q_1 i}  \nonumber  \\
    &+\bar{c}^{(q_1 q_3)}_{ik} \bar{P}_{q_2 j}+\bar{P}_{q_1 i} \bar{P}_{q_2 j} \bar{P}_{q_3 k},  \label{eq:tq1q2q3}
 \end{align}
respectively. The denominators $\bar{C}_H$'s are normalization constants given by
$\bar{C}_{\Lambda} = 1 - \bar{t}^{(ud)}_{ii}$,
$\bar{C}_{H_{112}} = 3 + \bar{t}^{(q_1 q_1)}_{ii} - 4 \bar{t}^{(q_1 q_2)}_{ii}$,
$\bar{C}_{\Sigma^0} = 3 + \bar{t}^{(ud)}_{ii} - 2 \bar{t}^{(us)}_{ii} - 2 \bar{t}^{(ds)}_{ii}$.

{\it Vector mesons:}
%
For vector mesons, we have a vector and a tensor polarization part. 
We present the results for tensor polarization here. 
\begin{align}
 S_{LL}=&\frac{1}{\bar{C}_V}\big(3\bar{t}^{(q_1 \bar{q}_2)}_{zz}-\bar{t}^{(q_1 \bar{q}_2)}_{ii}\big),  
 \nonumber
 \\
 S_{LT}^x=&\frac{2}{\bar{C}_V}  \big(\bar{t}^{(q_1 \bar{q}_2)}_{xz}+\bar{t}^{(q_1 \bar{q}_2)}_{zx}\big),  
 \nonumber
 \\
 S_{LT}^y=&\frac{2}{\bar{C}_V}  \big(\bar{t}^{(q_1 \bar{q}_2)}_{yz}+\bar{t}^{(q_1 \bar{q}_2)}_{zy}\big),      
 \nonumber
 \\
 S_{TT}^{xy}=&\frac{2}{\bar{C}_V}  \big(\bar{t}^{(q_1 \bar{q}_2)}_{xy}+\bar{t}^{(q_1 \bar{q}_2)}_{yx}\big),   
 \nonumber
 \\
 S_{TT}^{xx}=&\frac{2}{\bar{C}_V} \bigl(\bar{t}^{(q_1 \bar{q}_2)}_{xx}-\bar{t}^{(q_1 \bar{q}_2)}_{yy}\bigr),  \label{eq:Vpol}
 \end{align}
 where the normalization constant is $\bar{C}_V=3+ \bar{t}_{ii}^{(q_1\bar q_2)}$.  
 We see clearly that they are determined by the local spin correlation between $q_1$ and $\bar{q}_2$. 

{\it {Spin-3/2 hyperons:}} 
For  $J^P=(3/2)^+$ hyperons, the complete results are given in~\cite{Zhang:2024hyq}.
We only show the longitudinal components for comparisons. 
\begin{align}
    S_L=&\frac{1}{2 \bar{C}_3} \big(   
     5\sum_{j=1}^3  \bar{P}_{q_j z}+\bar{t}^{ \{ q_1 q_2 q_3\}}_{zii} \big ),
     \nonumber
     \\ 
    S_{LL}=& \frac{1}{\bar{C}_3}
     \big[(3\bar{t}^{(q_1 q_2)}_{zz}-\bar{t}^{(q_1 q_2)}_{ii})+c(123) \big],   
     \nonumber
     \\
    S_{LLL}=& \frac{9}{10 \bar{C}_3} \big(5\bar{t}^{(q_1 q_2 q_3)}_{zzz}-\bar{t}^{\{ q_1 q_2 q_3\} }_{zii} \big),
    \label{eq:slll}
\end{align}
where $\bar{C}_3=3+\bar{t}^{(q_1 q_2)}_{ii}+\bar{t}^{(q_2 q_3)}_{ii}+\bar{t}^{(q_3 q_1)}_{ii}$; 
$c(123)$ represents the cyclic exchange terms and 
$\bar{t}^{ \{ q_1q_2q_3\} }_{ijk}=\bar{t}^{(q_1q_2q_3)}_{ijk}+\bar{t}^{(q_3q_1q_2)}_{ijk}+\bar{t}^{(q_2q_3 q_1)}_{ijk}. $
We see clearly that while $S_L$ is determined by the quark spin polarizations and correlations, 
the second rank and third rank tensor polarizations $S_{LL}$ and $S_{LLL}$  are mainly determined by the local two and three quark spin correlations.  

\subsection{Hadron-hadron spin correlations} \label{sec:hhsc}

For hadron-hadron spin correlations, we present results for hyperon-hyperon ($HH$), hyperon-anti-hyperon ($H\bar H$), hyperon-vector meson ($HV$) and vector meson-vector meson ($VV$) respectively. 
For vector meson related spin correlations, we focus on the tensor polarization part since it can be measured via angular distributions of two body decays. 
Also we focus on longitudinal polarizations that can be measured via polar angle distributions of the decay products.

{\it $H\bar H$ and $HH$ spin correlations}: 
We take $\Lambda\bar{\Lambda}$ as an example and present the results here. It is given by
\begin{align}
    &\bar{t}^{\Lambda \bar{\Lambda}}_{ij}=\frac{1}{\bar{C}_{\Lambda \bar{\Lambda}} }
    \bigl[\bar{t}^{(s \bar{s})}_{ij}-\bar{t}^{(s \bar{u} \bar{d} \bar{s})}_{illj}-\bar{t}^{(uds \bar{s})}_{llij}+\bar{t}^{(uds \bar{u} \bar{d} \bar{s})}_{llikkj} \bigr],
    \label{eq:lambda and lambdabar} \\
   & \bar{C}_{\Lambda \bar{\Lambda}}=1-\bar{t}^{(ud)}_{ll}-\bar{t}^{(\bar{u} \bar{d})}_{ll}+\bar{t}^{(ud \bar{u} \bar{d})}_{llkk}.
    \label{eq:lambda constant}
\end{align}
Here $\bar{t}^{(q_1 q_2)}_{ij}$ and $ \bar{t}^{(q_1 q_2 q_3)}_{ijk}$
are contributions from two and three quark spin correlations given in Eqs.(\ref{eq:tq1q2}) and (\ref{eq:tq1q2q3}) respectively.  
Those from more than three quarks are defined by extending them in a straight forward way 
and are given in appendix~\ref{appendix:complete combination result} 
where we also present results for other hyperons and/or anti-hyperons. 

We emphasize that spin correlations of $H\bar H$ are determined by long-range quark-anti-quark spin correlations. 
We see that if we take long range spin correlation as zero (i.e., $\bar c_{ij}^{(q\bar q)}=0$) , we obtain $\bar{c}^{\Lambda \bar{\Lambda}}_{ij}=0$ and $\bar{C}_{\Lambda \bar{\Lambda}}=\bar{C}_{\Lambda}\bar{C}_{\bar\Lambda}$.

{\it $HV$ spin correlations}: 
Here we take $\Lambda V$ as an example and consider $u+d+s \to \Lambda$ and $q_4 +\bar{q}_5 \to V$. 
There are totally 24 independent spin correlation components.
We present only those relevant to the longitudinal tensor polarization $S_{LL}$ of the vector meson that can be studied most easily in experiments. 
The results are
 \begin{align}
    &  \bar{t}^{ \Lambda V}_{z,LL}=
        - \frac{1}{ \bar{C}_{ \Lambda V}} \bigr[\bar{t}^{( sq_4 \bar{q}_5)}_{zii}-3\bar{t}^{( sq_4 \bar{q}_5)}_{zzz}
        \nonumber      \\
    &~~~~~~~~~~~~~~~~~ -\bar{t}^{(u d s q_4 \bar{q}_5 )}_{iizjj}+3\bar{t}^{(u d s q_4 \bar{q}_5 )}_{iiz zz} \bigr],
        \label{eq:lambdaV} \\
 & \bar{C}_{\Lambda V}=3+\bar{t}^{(q_4 \bar{q}_5)}_{ii}-3\bar{t}^{(ud)}_{ii}
        -\bar{t}^{(ud q_4 \bar{q}_5 )}_{iijj}.
        \label{eq:lambdaV norm}
\end{align}

Results for other $HV$ combinations are given in Appendix~\ref{appendix:complete combination result}. 
We see that the spin correlations between $\Lambda$ and a vector meson can serve as a probe to three-quark (anti-quark) spin correlations. 
However, in contrast to spin-$3/2$ hyperons, where three quarks are all located in one hyperon thus called local quark spin correlations, 
those three-quark (anti-quark) spin correlations probed by $\bar{t}^{ \Lambda V}_{z,LL}$ are long range spin correlations. 
It is also easy to show that $\bar{c}^{ \Lambda V}_{z,LL}=0$ and $\bar{C}_{\Lambda V}=\bar{C}_{\Lambda}\bar{C}_{V}$ if we put the  long range quark (anti-quark) spin correlations as zero. 

{\it $VV$ spin correlations}: 
Here, we consider $q_1+\bar{q}_2 \to V_1$ and $q_3+\bar{q}_4 \to V_2$ and focus on $S_{LL}-S_{LL}$ correlations.  We obtain
    \begin{align}
        \bar{t}^{V_1 V_2}_{LL,LL}&=\frac{1}{ \bar{C}_{V_1 V_2}}
        \bigl[ \bar{t}^{(q_1 \bar{q}_2 q_3 \bar{q}_4)}_{iijj}-3\bar{t}^{(q_1 \bar{q}_2 q_3 \bar{q}_4)}_{iizz} 
        \nonumber
        \\
        &~~~~~~~~~~~~~~~~ -3(\bar{t}^{(q_1 \bar{q}_2 q_3 \bar{q}_4)}_{zzii}-3\bar{t}^{(q_1 \bar{q}_2 q_3 \bar{q}_4)}_{zzzz}) \bigr],
        \label{eq:VV} \\
    \bar{C}_{V_1 V_2}&=9+3 \bar{t}^{(q_1 \bar{q}_2)}_{ii}+3\bar{t}^{(q_3 \bar{q}_4)}_{ii} +\bar{t}^{(q_1 \bar{q}_2 q_3 \bar{q}_4)}_{iijj}.
    \label{eq:VV norm}
\end{align}
The complete results are given in Appendix~\ref{appendix:complete combination result}.
We see that $\bar{t}^{V_1 V_2}_{LL,LL}$ depends on the local $q_1\bar q_2$ and $q_3\bar q_4$ spin correlations and long range spin correlations between quark (anti-quark) in $V_1$ and that in $V_2$. 
It is also clear that if we put such  long range  spin correlations as zero we obtain $\bar{c}^{V_1 V_2}_{LL,LL}=0$ and $\bar{C}_{V_1 V_2}= \bar{C}_{V_1} \bar{C}_{V_2}$.

\subsection{Discussions}

The results presented above are for the general case and are quite lengthy. 
To see the physical consequences more clearly, we discuss a number of extreme cases and present the simplified results in the following of this section.

First of all, besides others, we consider only $\bar P_{qz}=\bar{P}_q$ while $\bar P_{qx}=\bar P_{qy}=0$, where the quark polarization 
is aligned to the normal direction of the reaction plane~\cite{Liang:2004ph,Liang:2004xn,Gao:2007bc} and we consider no flavor dependence. 
This is generally expected if the global polarization of QGP is purely due to the initial orbital angular momentum and QCD interactions 
and the QGP reaches  a local equilibrium before hadronization.
We consider also only spin correlations up to three quarks and/or anti-quarks. 
In this case, the results are simplified as
\begin{align}
P_{Hj}=&\bar{P}_q \delta_{zj}+\frac{\delta \rho_{Hj}^{(1)}}{\bar C_{H}^{(1)}},     \label{eq:PHsimple} \\
S_{LL}^V=&  \frac{1}{\bar{C}_V^{(1)}}\big(3\bar{c}^{(q_1 \bar{q}_2)}_{zz}-\bar{c}^{(q_1 \bar{q}_2)}_{ii}+2 \bar{P}_q^2 \big),     \label{eq:SVLLsimple}\\
S^{Va}_{LT}=&\frac{2}{\bar{C}_V^{(1)}}  \big(\bar{c}^{(q_1 \bar{q}_2)}_{a z}+\bar{c}^{(q_1 \bar{q}_2)}_{z a}\big),    \label{eq:SVLTsimple}\\
S^{Vxy}_{TT}=&\frac{2}{\bar{C}_V^{(1)}}  \big(\bar{c}^{(q_1 \bar{q}_2)}_{xy}+\bar{c}^{(q_1 \bar{q}_2)}_{yx}\big),  \label{eq:SVTTxysimple}\\
S^{Vxx}_{TT}=&\frac{2}{\bar{C}_V^{(1)}} \bigl(\bar{c}^{(q_1 \bar{q}_2)}_{xx}-\bar{c}^{(q_1 \bar{q}_2)}_{yy}\bigr) ,  \label{eq:SVTTxxsimple} \\
S_L^{H_3}=& \frac{5}{2} \bar{P}_q+ \frac{1}{2 \bar{C}_3^{(1)}}
\bigl\{ [\bar{c}^{(q_1 q_2 q_3)}_{zii}+2\bar{P}_q (\bar{c}^{(q_1 q_2)}_{zz}
\nonumber
\\
&-2\bar{c}^{(q_1 q_2)}_{ii} ) +c(123)]-12 \bar{P}_q^3 \bigr\},
\\
S_{LL}^{H_3}=& \frac{1}{\bar{C}_3^{(1)}}
     \bigl[(3\bar{c}^{(q_1 q_2)}_{zz}-\bar{c}^{(q_1 q_2)}_{ii} +c(123) )+6\bar{P}_q^2\big],      \\
S_{LLL}^{H_3}=&\frac{9}{10 \bar{C}_3^{(1)}} \bigl\{
2 \bar{c}^{(q_1 q_2 q_3)}_{zzz}+2 \bar{P}_q^3-
\bigl[ \bar{c}^{(q_1 q_2 q_3)}_{zxx}+\bar{c}^{(q_1 q_2 q_3)}_{zyy}
\nonumber
\\
&+\bar{P}_q (\bar{c}^{(q_1 q_2)}_{ii} -3 \bar{c}^{(q_1 q_2)}_{zz})  +c(123)\bigr]
\bigr\},    \label{eq:omega case I} \\
\bar{c}_{zz}^{\Lambda \bar \Lambda}=&\bar{c}^{(s \bar{s})}_{zz}-\frac{\bar{P}_q}{\bar{C}^{(1)}_{\Lambda}}
\bigl[ \bar{P}_q (\bar{c}^{(d \bar{s})}_{zz}+\bar{c}^{(u \bar{s})}_{zz})+\bar{c}^{(ud \bar{s})}_{iiz}
\nonumber
\\
&+\bar{c}^{(us \bar{s})}_{zzz}
+\bar{c}^{(d s \bar{s})}_{zzz}
\bigr]-\frac{\bar{P}_q}{\bar{C}_{\bar{\Lambda}}^{(1)}}
\bigl[  \bar{P}_q
(\bar{c}^{(\bar{d}s)}_{zz}+\bar{c}^{(\bar{u} s)}_{zz})
\nonumber
\\
&+\bar{c}^{(\bar{u} \bar{d}s)}_{iiz}+\bar{c}^{(\bar{u} s \bar{s})}_{zzz}+\bar{c}^{(\bar{d} s \bar{s})}_{zzz}
\bigr],
\label{eq: } 
\\
\bar{c}^{\Lambda \bar{H}_{112}}_{zz}=&\frac{1}{3}
(4 \bar{c}^{(s \bar{q}_1)}_{zz}-\bar{c}^{(s \bar{q}_2)}_{zz})-\frac{\bar{P}_q}{3\bar{C}_{\Lambda}^{(1)}}
\bigl[ ( \bar{P}_q ( 4 \bar{c}^{(u \bar{q}_1)}_{zz}- \bar{c}^{(u \bar{q}_2)}_{zz})
\nonumber
\\
&+4 \bar{c}^{(us \bar{q}_1)}_{zzz}-\bar{c}^{(us \bar{q}_2)}_{zzz}  +u \leftrightarrow d)+4\bar{c}^{(ud \bar{q}_1)}_{iiz}-\bar{c}^{(ud \bar{q}_2)}_{iiz}  \bigr]
\nonumber
\\
&- \frac{\bar{P}_q}{ \bar{C}_{\bar{H}_{112}}^{(1)}}
\bigl[ 2\bar{P}_q(2 \bar{c}^{(s \bar{q}_2)}_{zz}+\bar{c}^{(s \bar{q}_1)}_{zz})
+4\bar{c}^{(s \bar{q}_1 \bar{q}_1)}_{zzz}
\nonumber
\\
&-\bar{c}^{(s \bar{q}_1 \bar{q}_1)}_{zii}
+2(\bar{c}^{(s \bar{q}_1 \bar{q}_2)}_{zzz}+2 \bar{c}^{(s \bar{q}_1 \bar{q}_2)}_{zii})
\bigr],
\\
\bar{c}^{\Lambda \Lambda}_{zz} =& \bar{c}^{(ss)}_{zz}-\frac{2 \bar{P}_q}{\bar{C}^{(1)}_{\Lambda}}
\bigl[\bar{P}_q(\bar{c}^{(ds)}_{zz}+\bar{c}^{(us)}_{zz})+\bar{c}^{(uds)}_{iiz}
\nonumber
\\
&+\bar{c}^{(uss)}_{zzz}+\bar{c}^{(dss)}_{zzz}
\bigr],
\\
\bar{c}^{\Lambda H_{112}}_{zz}=&\frac{1}{3}
(4 \bar{c}^{(s {q}_1)}_{zz}-\bar{c}^{(s {q}_2)}_{zz})-\frac{\bar{P}_q}{3\bar{C}_{\Lambda}^{(1)}}
\bigl[ ( \bar{P}_q ( 4 \bar{c}^{(u {q}_1)}_{zz}- \bar{c}^{(u {q}_2)}_{zz})
\nonumber
\\
&+4 \bar{c}^{(us {q}_1)}_{zzz}-\bar{c}^{(us {q}_2)}_{zzz}  +u \leftrightarrow d)+4\bar{c}^{(ud {q}_1)}_{iiz}-\bar{c}^{(ud {q}_2)}_{iiz}  \bigr]
\nonumber
\\
&- \frac{\bar{P}_q}{ \bar{C}_{{H}_{112}}^{(1)}}
\bigl[ 2\bar{P}_q(2 \bar{c}^{(s {q}_2)}_{zz}+\bar{c}^{(s {q}_1)}_{zz})
+4\bar{c}^{(s {q}_1 {q}_1)}_{zzz}
\nonumber
\\
&-\bar{c}^{(s {q}_1 {q}_1)}_{zii}
+2(\bar{c}^{(s {q}_1 {q}_2)}_{zzz}+2 \bar{c}^{(s {q}_1 {q}_2)}_{zii})
\bigr],
\\
\bar{c}_{z,LL}^{\Lambda V} =&  \frac{1}{\bar{C}_{\Lambda V}^{(1)}} \bigl[ 3\bar{c}^{(s q_4 \bar{q}_5)}_{zzz}-\bar{c}^{(s q_4 \bar{q}_5)}_{zii} 
\nonumber
\\
&\qquad \quad +2\bar{P}_q (\bar{c}^{(s q_4)}_{zz}+\bar{c}^{(s \bar{q}_5)}_{zz}) \bigr],  \label{eq: } \\
\bar{c}_{LL,LL}^{V_1 V_2} =& \frac{2}{\bar{C}_{V_1 V_2}^{(1)}}
\bigl\{2 \bar{P}_q^2 \bigl[\bar{c}^{(q_1 q_3)}_{zz} 
+
\bar{c}^{(\bar{q}_2 \bar{q}_4)}_{zz} 
    +\bar{c}^{(q_1 \bar{q}_4)}_{zz}
    \nonumber
    \\
    &+\bar{c}^{(\bar{q}_2 q_3)}_{zz}
    \bigr]
    + \bar{P}_q \bigl[ 
   (3\bar{c}^{(q_1 \bar{q}_2 q_3)}_{zzz}-\bar{c}^{(q_1 \bar{q}_2 q_3)}_{iiz})
   \nonumber
   \\
    &+(3\bar{c}^{(q_1 q_3 \bar{q}_4)}_{zzz}-\bar{c}^{(q_1 q_3 \bar{q}_4)}_{zii}    )
     + (3\bar{c}^{(q_1 \bar{q}_2 \bar{q}_4)}_{zzz}
   \nonumber
   \\
   &-\bar{c}^{(q_1 \bar{q}_2 \bar{q}_4)}_{iiz})  +(3 \bar{c}^{(\bar{q}_2 q_3 \bar{q}_4)}_{zzz}-\bar{c}^{(\bar{q}_2 q_3 \bar{q}_4)}_{zii}) \bigr]
    \bigr\},  
    \label{eq: ar} 
\end{align}
where
\begin{align}
    \delta\rho_{\Lambda j}^{(1)}=&-\bigl[   
\bar{P}_q (\bar{c}^{(ds)}_{zj}+\bar{c}^{(us)}_{zj}) +\bar{c}^{(uds)}_{iij}
\bigr],
\\
\delta \rho_{H_{112} j}^{(1)}=&\bar{c}^{(q_1 q_1 q_2)}_{iij}-4\bar{c}^{(q_1 q_1 q_2)}_{jii}
\nonumber
\\
&-2\bar{P}_q \bigl[2 \bar{c}^{(q_1 q_1)}_{jz}- (\bar{c}^{(q_1 q_2)}_{zj}-2 \bar{c}^{(q_1 q_2)}_{jz}) \bigr],
\\
\bar{C}^{(1)}_{\Lambda}=&
1-\bar{c}^{(ud)}_{ii}-\bar{P}_q^2,
\\
\bar{C}^{(1)}_{H_{112}}=&3+\bar{c}^{(q_1 q_1)}_{ii}-4 \bar{c}^{(q_1 q_2)}_{ii}-3\bar{P}_q^2,
\\
\bar{C}^{(1)}_V=&3+\bar{c}^{(q_1 \bar{q}_2)}_{ii}+\bar{P}_q^2,
\\
\bar{C}_3^{(1)}=&3 +\bar{c}^{(q_1 q_2)}_{ii}
+\bar{c}^{(q_1 q_3)}_{ii}
+\bar{c}^{(q_2 q_3)}_{ii}+3 \bar{P}_q^2,
\\
\bar{C}_{\Lambda V}^{(1)}
=& 3+\bar{c}^{(q_4 \bar{q}_5)}_{ii}-3\bar{c}^{(ud)}_{ii}-2\bar{P}_q^2,
\\
\bar{C}_{V_1 V_2}^{(1)}
    =& 9+3 \bar{c}^{(q_1 \bar{q}_2)}_{ii}+3 \bar{c}^{(q_3 \bar{q}_4)}_{ii}+6 \bar{P}_q^2.
\end{align}

Furthermore, we discuss these results in following extreme cases for quark spin correlations.

{\it Case 0:} We neglect all quark spin correlations, i.e. we take $\bar c^{(q_1q_2)}_{ij}=\bar c^{(q_1\bar q_2)}_{ij}=0$ 
and similarly for those of more than two quarks (anti-quarks). 
This is the case discussed by Liang and Wang in their original papers in 2005~\cite{Liang:2004ph,Liang:2004xn}.

In this case, we obtain the well-known results
\begin{align}
&P_{H j}=\bar{P}_q \delta_{zj},      \label{eq:case0PH} \\
&S_{LL}^V=\frac{2\bar{P}_q^2 }{3+\bar{P}_q^2 }, \\
&S^{V a}_{LT}= S^{Vx a}_{TT}=0, \\
&S_L^{H_3}=\frac{5}{2} \bar{P}_q -\frac{2 \bar{P}_q^3}{1+\bar{P}_q^2},  \\
&S_{LL}^{H_3}=\frac{2 \bar{P}_q^2}{1+ \bar{P}_q^2},      \\
&S_{LLL}^{H_3}=\frac{3 \bar{P}_q^3}{5 (1+ \bar{P}_q^2)} ,    \label{eq:case0SLLL} \\
&\bar{c}_{ij}^{H_1H_2} =\bar{c}_{ij}^{H_1\bar H_2}=\bar{c}_{ij}^{HV} =\bar{c}_{ij}^{V_1V_2} =0.  
\end{align}
 Here we use $H_3$ to denote hyperons of $J^P=(3/2)^+$ to distinguish it from those of $J^P=(1/2)^+$. 
 
 We see that hyperon polarizations including tensor polarizations have a series of simple relationships between each other 
 since they all are determined by the single quantity $\bar P_q$. 
 Significant deviations of experimental results and these simple relationships 
 may imply flavor dependence of quark polarizations and/or existences of strong quark spin correlations. 
  
{\it Case I:} We consider only local quark spin correlations without flavor dependence or off diagonal components,
i.e. we take $\bar c^{(q_1\bar q_2)}_{ij,(l)}=\bar c^{(q\bar q)}_{(l)}\delta_{ij}$,    
$\bar c^{(q_1q_2)}_{ij,(l)}=\bar c^{(\bar q_1\bar q_2)}_{ij,(l)}=\bar c^{(qq)}_{(l)}\delta_{ij}$, and similar for three quarks (anti-quarks).
while $\bar c^{(q_1\bar q_2)}_{ij,(L)}=\bar c^{(q_1q_2)}_{ij,(L)}=0$. 
Here we add a subscript $(l)$ for local and $(L)$ for long range quark spin correlations.

In this case, we obtain 
\begin{align}
&P_{H j}= \bar{P}_q \delta_{zj}- \frac{2 \bar{P}_q \delta_{zj} \bar{c}^{(qq)}_{(l)}  +3\bar{c}^{(qqq)}_{(l)} }{1-3\bar{c}^{(qq)}_{(l)}-\bar{P}_q^2},      \label{eq:case1PH} \\
&S_{LL}^V= \frac{2\bar{P}_q^2} {3+3\bar{c}^{(q \bar{q})}_{(l)}+\bar{P}_q^2},  \label{eq:case1SLL}
\\
&S^{V a}_{LT}=S^{Vx a}_{TT}=0, \\
&S_L^{H_3}=\frac{5}{2} \bar{P}_q  + \frac{ \bar{c}^{(qqq)}_{(l)} -10\bar{P}_q \bar{c}^{(qq)}_{(l)}  -4\bar{P}_q^3 } 
{2(1+3\bar{c}^{(qq)}_{(l)}+\bar{P}_q^2)}, \\
&S_{LL}^{H_3}= \frac{2\bar{P}_q^2}{1+3\bar{c}^{(qq)}_{(l)}+\bar{P}_q^2}, \\
&S_{LLL}^{H_3}=\frac{3}{5}\frac{\bar{c}^{(qqq)}_{(l)}+ \bar{P}_q^3}{1+3\bar{c}^{(qq)}_{(l)}+\bar{P}_q^2}, \label{eq:case1SLLL} \\
&\bar{c}_{zz}^{H_1H_2} =\bar{c}_{zz}^{H_1\bar H_2} =
\bar{c}_{z,LL}^{HV} = \bar{c}_{LL,LL}^{V_1 V_2} =0.  \label{eq:case1chh} 
\end{align}

We see clearly that local quark spin correlations lead to significant contributions to both vector and tensor polarizations of $J^P=(3/2)^+$ baryons. 
However, since we consider isotropic quark-antiquark spin correlations $\bar c^{(q_1\bar q_2)}_{xx,(l)}= \bar c^{(q_1\bar q_2)}_{yy,(l)}=\bar c^{(q_1\bar q_2)}_{zz,(l)}$, 
they cancel each other in the numerator of $S_{LL}^V$.

{\it Case II:} We consider only local quark spin correlations and assume no flavor independence but a difference between $z$ and $x$ or $y$ directions, i.e., 
$\bar{c}^{(q_1 \bar q_2)}_{xx/yy(l)} = \bar{c}^{(q\bar q)}_{\perp(l)}\not=\bar{c}^{(q\bar q)}_{zz(l)}$. 

In this case, we obtain 
\begin{align}
    &P_{Hj}= \bar{P}_q \delta_{zj}
    -\frac{2 \bar{P}_q \bar{c}^{(qq)}_{zz(l)}  \delta_{zj} +2 \bar{c}^{(qqq)}_{\perp(l)}+\bar{c}^{(qqq)}_{zzz(l)}}{1-\bar{c}^{(qq)}_{zz(l)}-2\bar{c}^{(qq)}_{\perp(l)}-\bar{P}_q^2},
    \label{eq: case2PH} \\
    &S^{V}_{LL}= \frac{2(\bar{c}^{(q \bar{q})}_{zz(l)}-\bar{c}^{(q \bar{q})}_{\perp(l)}+\bar{P}_q^2)}{3+\bar{c}^{(q \bar{q})}_{zz(l)}+2\bar{c}^{(q \bar{q})}_{\perp(l)}
    +\bar{P}_q^2}, \label{eq: case2SLL}
    \\
    &S^{V a}_{LT}= S^{V x a}_{TT}=0,
    \\
    &S^{H_3}_{L}= \frac{5}{2} \bar{P}_q
    +\frac{\bar{c}^{(qqq)}_{zzz(l)}-2 (\bar{c}^{(qq)}_{zz(l)}+4\bar{c}^{(qq)}_{\perp(l)})-4 \bar{P}_q^3   }{2(1+\bar{c}^{(qq)}_{zz(l)}+2 \bar{c}^{(qq)}_{\perp(l)}+\bar{P}_q^2)},
    \\
    &S^{H_3}_{LL}= \frac{2(\bar{c}^{(qq)}_{zz(l)}-\bar{c}^{(qq)}_{\perp(l)}+\bar{P}_q^2)}{1+\bar{c}^{(qq)}_{zz(l)}+2 \bar{c}^{(qq)}_{\perp(l)}+\bar{P}_q^2},
    \\
    &S^{H_3}_{LLL}= \frac{3}{5}
    \frac{\bar{c}^{(qqq)}_{zzz(l)}+3\bar{P}_q (\bar{c}^{(qq)}_{zz(l)}-\bar{c}^{(qq)}_{\perp(l)})
    + \bar{P}_q^3}{1+\bar{c}^{(qq)}_{zz(l)}+2 \bar{c}^{(qq)}_{\perp(l)}+\bar{P}_q^2},  \label{eq:case2SLLL}
    \\
& \bar{c}_{zz}^{H_1H_2}= \bar{c}_{zz}^{H_1\bar H_2}= \bar{c}_{z,LL}^{HV} =\bar{c}_{LL,LL}^{V_1 V_2} =0.
\label{eq: case2cLLLL} 
\end{align}

From Eqs.(\ref{eq: case2PH}-\ref{eq:case2SLLL}), we see that tensor polarizations are quite sensitive to the difference between $\bar{c}_{zz}$ and $\bar{c}_{\perp}$.
This difference might dominate $S_{LL}$ and $S_{LLL}$.

{\it Case III:} We consider both local and long range quark spin correlations and assume no flavor independence but a difference between $z$ and $x$ or $y$ directions, i.e., 
$\bar{c}^{(q_1 \bar q_2)}_{xx/yy(l)} = \bar{c}^{(q\bar q)}_{\perp(l)}\not=\bar{c}^{(q\bar q)}_{zz(l)}$, and similar for $\bar{c}^{(q_1 \bar q_2)}_{xx/yy/zz(L)}$. 

In this case, the hadron spin polarizations remain the same as those in case II. 
For hadron spin correlations, we obtain 
\begin{align}
& \bar{c}_{zz}^{H_1H_2}= \bar{c}^{(qq)}_{zz(L)}-2 \bar{P}_q \frac{ 2 \bar{P}_q \bar{c}^{(qq)}_{zz(L)}+3\bar{c}^{(qqq)}_{zzz(lL)} }{1-\bar{c}^{(qq)}_{zz(l)}-2\bar{c}^{(qq)}_{\perp(l)}-\bar{P}_q^2},   \label{eq: case3cHH} \\
& \bar{c}_{zz}^{H_1\bar H_2}= \bar{c}^{(q \bar{q})}_{zz(L)} 
-2\bar{P}_q \frac{  2\bar{P}_q \bar{c}^{(q \bar{q})}_{zz(L)}+3\bar{c}^{(qq\bar{q})}_{zzz(lL)}  }{1-\bar{c}^{(qq)}_{zz(l)}-2\bar{c}^{(qq)}_{\perp(l)}-\bar{P}_q^2} ,    \\
&\bar{c}_{z,LL}^{HV} =
\frac{2} {\bar{C}_{HV}} \bigl[ \bar{c}^{(qq \bar{q})}_{zzz(lL)}+ \bar{P}_q (\bar{c}^{(qq)}_{zz(L)}+\bar{c}^{(q \bar{q})}_{zz(L)}) \bigr] , \\
&\bar{c}_{LL,LL}^{V_1 V_2} =\frac{8\bar{P}_q  }{\bar{C}_{V_1 V_2 }}
\bigl[ 2\bar{c}^{(qq \bar{q})}_{zzz(lL)} + \bar{P}_q(\bar{c}^{(qq)}_{zz(L)} + \bar{c}^{(q \bar{q})}_{zz(L)} ) \bigr],
\label{eq: case3cLLLL} 
\end{align}
and the normalization constants are given by
\begin{align}
     \bar{C}_{HV}=&3+2(\bar{c}^{(q \bar{q})}_{\perp(l)}-3\bar{c}^{(qq)}_{\perp(l)})
     \nonumber
     \\
    &+\bar{c}^{(q \bar{q})}_{zz(l)}-3 \bar{c}^{(qq)}_{zz(l)}-2\bar{P}_q^2,
    \\
    \bar{C}_{V_1 V_2 }=&3(3+ 4\bar{c}^{(q \bar{q})}_{\perp(l)}+2\bar{c}^{(q \bar{q})}_{zz(l)}+2 \bar{P}_q^2).
    \label{eq:case3cLLLL}
\end{align}

Here $\bar{c}^{(qqq)}_{(lL)}$ represents a mixed local and long range quark spin correlation, i.e., two of them are local (in one hadron) and the third is long range (in another hadron). 
It is given by $\bar{c}^{(qqq)}_{(lL)} = \langle h_1h_2 | c^{(qqq)}| h_1h_2 \rangle$, where two of the quarks are e.g. in $h_1$ while the third is in $h_2$.
 
From Eqs.(\ref{eq: case3cHH}-\ref{eq:case3cLLLL}), it is evident that the hadron spin correlations are none zero only if long range quark correlations are taken into account. 
We see also that for longitudinal hadron spin correlations, transverse quark spin correlation $c_\perp$ appears only in the denominator, i.e., in the normalization. 
This shows that hadron spin correlations probe more significantly quark spin correlations in the same direction.

{\it Case IV:} We consider the case that only induced quark spin correlations exist. 
As discussed in \cite{Lv:2024uev}, we have only local quark spin correlations if we do not consider the overlap of the wave functions of hadrons. 
If we consider no flavor dependence or off diagonal components, 
this just corresponds to case II with replacements,   
$\bar c^{(qq)}_{zz(l)}=\langle P_{qz}P_{qz}\rangle_l-\bar{P}_q^2$, $\bar c^{(qq)}_{\perp(l)}=\langle P_{qx}P_{qx}\rangle_l$. 
Here $\langle P_{qz}P_{qz}\rangle_l\equiv\langle h|P_{q_1z}(\alpha_{q_1})P_{q_2z}(\alpha_{q_2})|h\rangle$ 
and similar for other and those for $q\bar q$.
Then we have 
\begin{align}
&P_{H j}=\bar{P}_q \delta_{zj} \bigl[1  -\frac{2}{\bar{C}_{H}}(\langle P_{qz}P_{qz} \rangle -\bar{P}_q^2) \bigr] \nonumber\\
&~~~~~~ -\frac{1}{\bar{C}_{H}} \bigl[2\langle P_{qx}P_{qx} P_{qx} \rangle
+\langle P_{qz}P_{qz} P_{qz} \rangle +\nonumber\\
&~~~~~~ -3\bar{P}_q  \langle P_{qz}P_{qz}  \rangle +2\bar{P}_q^3 \bigr],     \\
&S_{LL}^V= \frac{2}{\bar{C}_V}
\bigl[  \langle P_{qz }P_{\bar{q}z} \rangle
-\langle P_{qx} P_{\bar{q} x} \rangle\bigr], \\
&S^{V a}_{LT}=S^{Vxa}_{TT}=0,\\
&S_L^{H_3}=\frac{5}{2} \bar{P}_q+ \frac{1}{2\bar{C}_3}
    \bigl[   
    \langle P_{qz}P_{qz} P_{qz} \rangle \nonumber\\
&~~~~~~ -\bar{P}_q (5 \langle P_{qz}P_{qz} \rangle +8\langle P_{qx}P_{qx} \rangle ) \bigr],       \\
&S_{LL}^{H_3}= \frac{4}{\bar{C}_3}
   \bigl[  \langle P_{qz}P_{qz} \rangle -\langle P_{qx}P_{qx}  \rangle  \rangle
   \bigr],  \\
&S_{LLL}^{H_3}= \frac{3}{5 \bar{C}_3} 
( \langle P_{qz}P_{qz} P_{qz} \rangle-3\bar{P}_q \langle P_{qx}P_{qx}  \rangle ),    
\label{eq: case V} \\
&\bar{c}_{zz}^{H_1H_2} =\bar{c}_{zz}^{H_1\bar H_2} =\bar{c}_{z,LL}^{HV} = \bar{c}_{LL,LL}^{V_1 V_2} =0.  \label{eq:case4chh} 
\end{align}
and the normalization constants are 
\begin{align}
    &\bar{C}_{H}= 1 -2\langle P_{qx}P_{qx}  \rangle  -\langle P_{qz}P_{qz} \rangle,
    \\
   & \bar{C}_V=3+ 2\langle P_{qx}P_{\bar{q}x} \rangle+\langle P_{qz} P_{\bar{q}z} \rangle,
    \\
    &\bar{C}_3=1+2\langle P_{qx}P_{qx} \rangle  +\langle P_{qz}P_{qz} \rangle.
\end{align}

 We see that in this case hadron tensor polarizations such as $S_{LL}^V$ and $S_{LL}^{H_3}$ are mainly come from 
 the difference between $\langle P_{qz }P_{\bar{q}z} \rangle$ and $\langle P_{qx} P_{\bar{q} x} \rangle$. 
 We also note that the hadron spin correlations vanish for hadrons at given $\alpha_{h_1}$ and $\alpha_{h_2}$.  
This only possible origin is due to average in phase space over $(\alpha_{h_1}, \alpha_{h_2})$. 
Ref.~\cite{Sheng:2026hsc} shows in fact such an example where a correlation in space-time in a Gaussian form is introduced 
that leads to induced spin correlations between $\Lambda$ and/or $\bar\Lambda$ hyperons.

\section{Summary and outlook}\label{sec:summary}

  In this paper, we extend the study of global polarization and spin correlations in the QGP developed in~\cite{Lv:2024uev} 
  and present a systematic and complete set of results for final-state hadrons with different spins produced in relativistic heavy ion collisions. 
 The studies presented include spin-1/2 hyperons, spin-1 vector mesons, and spin-3/2 hyperons.
 We study their spin polarizations and spin correlations between them, including hyperon–hyperon, hyperon–vector meson and vector meson–vector meson.  
 We unify the definitions, summarize the formulae to measure them via decay processes, present the complete results for expressions of them in terms of the spin density matrix elements, and those in terms of quark spin polarizations and correlations in the quark combination mechanism. 
 We also make discussions of the results in simplified extreme cases. 
  
 These results show in particular that, as well known, vector polarizations of spin-1/2 hyperons can be measured by their weak decays, 
 tensor polarizations of vector mesons can be measured by their two body decays into two spinless mesons,  
 and many components of polarizations of spin-3/2 baryons can also be measured by their two body decays or successive two body decays. 
 The spin correlations between the corresponding polarization components of these hadrons can also be measured via these decay processes. 
 
 The results obtained in the quark mechanism show the following characteristics: 
 (1) Spin vector polarizations of hadrons such as polarizations of spin 1/2-hyperons and vector polarizations of spin-1 and spin-3/2 hadrons 
 are determined mainly by the average values of quark spin polarizations. There are however also influences from quark spin correlations. 
The tensor polarizations including vector meson spin alignments and rank-2 and 3 tensor polarizations  of spin-3/2 baryons are 
mainly determined by the corresponding quark spin correlations. 
 
 (2) There exist a series of simple relationships between different components of hadron spin correlations if quark spin correlations are neglected. 
 Such simple relationships can serve as tests of existences of quark spin correlations in the quark system produced in heavy ion collisions. 
 
(3) Vector meson spin alignments and rank-2 tensor polarizations of spin-3/2 baryons are quite sensitive to the difference of 
quark spin correlations in longitudinal and transverse directions. 

(4) Quark spin correlations include genuine and induced parts that cannot be separated from each other. 
Hadron-hadron spin correlations may include induced parts at the hadron level.  
 
 We should also mention that the study in this paper is based on the framework formulated in \cite{Lv:2024uev}. 
 It is envisaged that in heavy ion collisions a quark matter system is produced and hadrons are produced via quark combination mechanism. 
 The combination mechanism used is non-relativistic so that spin and other degree of freedom are factorized. 
 Further studies including relativistic effects and/or other hadronization mechanisms are needed and are underway.

\section*{Acknowledgments}
This work was supported in part by the National
Natural Science Foundation of China (Grant No.12375075 and No.12321005)  
and  Shandong Provincial Natural Science Foundation (Grant No.ZFJH202303 and Youth Program No.ZR2025QC1472).

\appendix
\begin{widetext}

\section{Angular distributions of decay products of $h_1h_2$}\label{appendix:decay}

 In high-energy reactions, polarization of produced particles is primarily determined by measuring the angular distributions of the decay products from the two body decay process $A \to 1+2$ in the rest frame of $A$.
The corresponding formula is derived using symmetry properties and conservation laws in the decay process and is given by~\cite{Lee:1957qs}
\begin{align}
W_A&(\theta,\varphi)=N\sum_{\lambda_i;m_A,m_A^\prime} \rho^A_{m_Am_A^\prime} |H_A(\lambda_1,\lambda_2)|^2   e^{i(m_A-m^\prime_A)\varphi} d_{m_A\lambda_{12}}^{j_A*}(\theta) d_{m_A^\prime\lambda_{12}}^{j_A}(\theta), 
\label{eq:A1}
\end{align}
where $j_A$, $m_A$ and $\rho^A_{m_Am_A^\prime}$ are spin, its third component and the element of the spin density matrix of $A$; 
$\lambda_1$ and $\lambda_2$ are helicities of $1$ and $2$, $\lambda_{12}=\lambda_1-\lambda_2$, $H_A(\lambda_1,\lambda_2)$ is the helicity amplitude of $A\to 1+2$; 
$d_{m^{\prime} m}^{j}(\theta)=\langle jm^{\prime}|e^{-i\theta \hat J_y}|jm\rangle$ is the element of the Wigner rotation matrix, and $N$ is a normalization constant. 
Results for hadrons with different spins such as spin-1/2 hyperons, vector mesons and spin-3/2 hyperons, even including successive decays  
can be found e.g., in~\cite{Lee:1957qs,Jacob:1959at,Rose1957,Chung:1971ri,Doncel:1972ez,Zhang:2023box}.  
We rewrite the result for $V\to M_1M_2$ ( two spin-0 mesons) in terms of $P_V^{(j)}$ and obtain

\begin{align}
W_V(\theta,\varphi)=&\frac{1}{4\pi}\Bigl[ 1+S_{LL}(1 - 3\cos^2\theta) -\frac{3}{2}\sin2\theta(S_{LT}^x\cos\varphi+S_{LT}^y\sin\varphi) -\frac{3}{2}\sin^2\theta(S_{TT}^{xx}\cos2\varphi+S_{TT}^{xy}\sin2\varphi)\Bigr].
\label{eq:A2}
\end{align}
it is evident from Eq.~(\ref{eq:A2}) that only the tensor polarization components contribute to the angular distribution.

The derivation can be straightforwardly extended to two hadron decays $h_1\to 1+1^\prime$ and $h_2\to 2+2^\prime$ if we assume that the two hadrons have a joint spin density matrix $\hat\rho^{h_1h_2}$ 
but decay independently. 
In this case, we have 

\begin{align}
W_{h_1h_2}&(\theta_1,\varphi_1;\theta_2,\varphi_2)=N\sum_{\lambda_i;m_{h_j}m_{h_j}^\prime} \rho^{h_1h_2}_{m_{h_1}m_{h_2}m_{h_1}^\prime m_{h_2}^\prime}  |H_{h_1}(\lambda_1,\lambda_2)|^2 |H_{h_2}(\lambda_3,\lambda_4)|^2   \nonumber\\
&\times e^{i(m_{h_1}-m^\prime_{h_1})\varphi_1} 
e^{i(m_{h_2}-m^\prime_{h_2})\varphi_2} d_{m_{h_1}\lambda_{11^\prime}}^{j_{h_1}*}(\theta_1) d_{m_{h_1}^\prime\lambda_{11^\prime}}^{j_{h_1}}(\theta_1)  d_{m_{h_2}\lambda_{22^\prime}}^{j_{h_2}*}(\theta_2) d_{m_{h_2}^\prime\lambda_{22^\prime}}^{j_{h_2}}(\theta_2).
\label{eq:A3}
\end{align}
Specifically, for the case where $h_1$ and $h_2$ are hyperons ($H_1$ and $H_2$) decaying via the weak channels $H_1\to N_1\pi_1$ and $H_2\to N_2\pi_2$, the joint angular distribution is given by
\begin{align}
W_{H_1 H_2}(\theta_1,\varphi_1;\theta_2,\varphi_2)
&= \frac{1}{(4\pi)^2} \Bigl\{
    1 + \alpha_{H_1} ( P_{H_1 x} \sin\theta_1 \cos\varphi_1 + P_{H_1 y} \sin\theta_1 \sin\varphi_1 + P_{H_1 z} \cos\theta_1 )
    \nonumber \\
    &\qquad + \alpha_{H_2} ( P_{H_2 x} \sin\theta_2 \cos\varphi_2 + P_{H_2 y} \sin\theta_2 \sin\varphi_2 + P_{H_2 z} \cos\theta_2 )
    \nonumber \\
    &\qquad + \alpha_{H_1}\alpha_{H_2} \bigl[
        \sin\theta_1 \cos\varphi_1 ( t^{H_1 H_2}_{xx} \sin\theta_2 \cos\varphi_2
        + t^{H_1 H_2}_{xy} \sin\theta_2 \sin\varphi_2
        + t^{H_1 H_2}_{xz} \cos\theta_2 )
        \nonumber \\
        &\qquad+ \sin\theta_1 \sin\varphi_1 ( t^{H_1 H_2}_{yx} \sin\theta_2 \cos\varphi_2
        + t^{H_1 H_2}_{yy} \sin\theta_2 \sin\varphi_2
        + t^{H_1 H_2}_{yz} \cos\theta_2 )
        \nonumber \\
        &\qquad  + \cos\theta_1 ( t^{H_1 H_2}_{zx} \sin\theta_2 \cos\varphi_2
        + t^{H_1 H_2}_{zy} \sin\theta_2 \sin\varphi_2
        + t^{H_1 H_2}_{zz} \cos\theta_2 )
    \bigr]
\Bigr\}.
\label{eq:A4}
\end{align}
If we consider only longitudinal (i.e. along $z$-direction) polarization, Eq.~(\ref{eq:A4}) simplifies to
\begin{align}
W_{H_1H_2}&(\theta_1,\theta_2)=\frac{1}{4}\Bigl[ 1 +\alpha_{H_1} P_{H_1z}\cos\theta_1 +\alpha_{H_2} P_{H_2z}\cos\theta_2 +\alpha_{H_1}\alpha_{H_2} t^{H_1 H_2}_{zz} \cos\theta_1\cos\theta_2\Bigr] .
\label{eq:A5}
\end{align}
At present, the measurement of $\Lambda_1 \Lambda_2$
 correlations in pp collisions is performed using~\cite{Gong:2021bcp,Shen:2024buh,STAR:2025njp}
 \begin{align}
     \frac{1}{N} \frac{d N}{ d \cos{\theta^*}}= \frac{1}{2}
     \left[1+\alpha_{H_1} \alpha_{H_2} P_{\Lambda_1 \Lambda_2} \cos{\theta^*} \right],
     \label{eq:A6}
 \end{align}
where the observable $P_{\Lambda_1 \Lambda_2}$ is related to the correlation coefficients $t^{\Lambda_1 \Lambda_2}_{ij}$ via
\begin{align}
     P_{\Lambda_1 \Lambda_2}=&\frac{3}{\alpha_{H_1} \alpha_{H_2}} \langle \cos{\theta^*} \rangle
     =\frac{1}{3} \bigr( t^{\Lambda_1 \Lambda_2}_{xx}+t^{\Lambda_1 \Lambda_2}_{yy}+t^{\Lambda_1 \Lambda_2}_{zz} \bigr),
     \label{eq:A7}
\end{align}
this shows explicitly that Eq.(\ref{eq:A6}) probes the diagonal correlation terms of the $\Lambda_1 \Lambda_2$ system.
For the case that $h_1$ is a hyperon $H$ and $h_2$ is a vector meson $V$ where
$H$ decay through
$H\to N\pi$ and $V$ decays into two spin-0 mesons, the corresponding angular distribution is given by
\begin{align}
W_{HV}(\theta_1,\varphi_1;&\theta_2,\varphi_2)
= \frac{1}{(4 \pi)^2} \Bigl\{
   1 + S_{LL}(1-3 \cos^2{\theta_2})-\frac{3}{2} \sin{2 \theta_2} \bigl(S^x_{LT} \cos{\varphi_2} + S^y_{LT} \sin{\varphi_2}\bigr)
   \nonumber
   \\
   &
   -\frac{3}{2} \sin^2{\theta_2} \bigl(S^{xx}_{TT} \cos{2 \varphi_2} + S^{xy}_{TT} \sin{2 \varphi_2}\bigr)
   + \alpha_H \cos{\theta_1} \bigl[
       P_{H z} + {t}^{HV}_{z,LL} (1-3\cos^2{\theta_2})
       \nonumber \\
       & -\frac{3}{2} \sin{2 \theta_2} \bigl({t}^{HV}_{z,^x_{LT}} \cos{\varphi_2} + {t}^{HV}_{z,^y_{LT}} \sin{\varphi_2}\bigr)
       -\frac{3}{2} \sin^2{\theta_2} \bigl({t}^{HV}_{z,^{xx}_{TT}} \cos{2 \varphi_2} + {t}^{HV}_{z,^{xy}_{TT}} \sin{2 \varphi_2}\bigr)
       \bigr]
   \nonumber \\
   & + \alpha_H \sin{\theta_1}\cos{\varphi_1} \bigl[
       P_{H x} + {t}^{HV}_{x,LL} (1-3\cos^2 \theta_2)
       \nonumber \\
       & -\frac{3}{2} \sin{2\theta_2} \bigl({t}^{HV}_{x,^x_{LT}} \cos{\varphi_2} + {t}^{HV}_{x,^{y}_{LT}} \sin{\varphi_2}\bigr)
       -\frac{3}{2} \sin^2{\theta_2} \bigl({t}^{HV}_{x,^{xx}_{TT}} \cos{2\varphi_2} + {t}^{HV}_{x,^{xy}_{TT}} \sin{2\varphi_2}\bigr)
       \bigr]
   \nonumber \\
   & + \alpha_H \sin{\theta_1}\sin{\varphi_1} \bigl[
       P_{H y} + {t}^{HV}_{y,LL} (1-3\cos^2 \theta_2)
       \nonumber \\
       & -\frac{3}{2} \sin{2\theta_2} \bigl({t}^{HV}_{y,^x_{LT}} \cos{\varphi_2} + {t}^{HV}_{y,^y_{LT}} \sin{\varphi_2}\bigr)
       -\frac{3}{2} \sin^2{\theta_2} \bigl({t}^{HV}_{y,^{xx}_{TT}} \cos{2\varphi_2} + {t}^{HV}_{y,^{xy}_{TT}} \sin{2\varphi_2}\bigr)
       \bigr]
\Bigr\}.
\label{eq:A8}
\end{align}
Similarly, by considering only the longitudinal (i.e. along the z-axis) polarization, Eq.~(\ref{eq:A8}) is further simplified as
\begin{align}
W_{HV}(\theta_1,\theta_2)=\frac{1}{4}\Bigl[1+S_{LL}(1-3\cos^2{\theta_2})+\alpha_{H} P_{Hz}\cos{\theta_1} +\alpha_H t^{HV}_{z,LL} \cos{\theta_1}(1-3\cos^2{\theta_2})\Bigr].  
\end{align}
For the case that $h_1h_2=V_1V_2$ and each of them decays into two spin-0 mesons respectively, we obtain
\begin{align}
W_{V_1V_2}(\theta_1,\varphi_1;\theta_2,&\varphi_2) = \frac{1}{(4 \pi)^2} \Bigl\{
    1 + \bigl[ S_{LL}(V_1)(1-3 \cos^2{\theta_1})
    -\frac{3}{2}\sin{2\theta_1} \bigl(S^{x}_{LT}(V_1) \cos{\varphi_1} + S^{y}_{LT}(V_1) \sin{\varphi_1}\bigr)
    \nonumber \\
    &\qquad\qquad\qquad\qquad -\frac{3}{2}\sin^2{\theta_1} \bigl(S^{xx}_{TT}(V_1) \cos{2\varphi_1} + S^{xy}_{TT}(V_1) \sin{2\varphi_1}\bigr) \bigr]
    \nonumber \\
    &+\bigl[ S_{LL}(V_2)(1-3 \cos^2{\theta_2})
    -\frac{3}{2}\sin{2\theta_2} \bigl(S^{x}_{LT}(V_2) \cos{\varphi_2} + S^{y}_{LT}(V_2) \sin{\varphi_2}\bigr)
    \nonumber \\
    &\qquad\qquad\qquad\qquad -\frac{3}{2}\sin^2{\theta_2} \bigl(S^{xx}_{TT}(V_2) \cos{2\varphi_2} + S^{xy}_{TT}(V_2) \sin{2\varphi_2}\bigr) \bigr]
    \nonumber \\
    & + (1-3\cos^2{\theta_1}) \bigl[
          {t}^{V_1 V_2}_{LL,LL} (1-3\cos^2{\theta_2})
        -\frac{3}{2}\sin{2\theta_2} \bigl({t}^{V_1 V_2}_{LL,^x_{LT}} \cos{\varphi_2} + {t}^{V_1 V_2}_{LL,^y_{LT}} \sin{\varphi_2}\bigr)
        \nonumber \\
        &\qquad\qquad\qquad\qquad -\frac{3}{2}\sin^2{\theta_2} \bigl({t}^{V_1 V_2}_{LL,^{xx}_{TT}} \cos{2\varphi_2} + {t}^{V_1 V_2}_{LL,^{xy}_{TT}} \sin{2 \varphi_2}\bigr)
    \bigr]
    \nonumber \\
    & -\frac{3}{2}\sin{2\theta_1} \cos{\varphi_1} \bigl[
         {t}^{V_1 V_2}_{^x_{LT},LL} (1-3\cos^2{\theta_2})
        -\frac{3}{2}\sin{2\theta_2} \bigl({t}^{V_1 V_2}_{^x_{LT},^x_{LT}} \cos{\varphi_2} + {t}^{V_1 V_2}_{^x_{LT},^y_{LT}} \sin{\varphi_2}\bigr)
        \nonumber \\
    &\qquad\qquad\qquad\qquad -\frac{3}{2}\sin^2{\theta_2} \bigl({t}^{V_1 V_2}_{^x_{LT},^{xx}_{TT}} \cos{2\varphi_2} + {t}^{V_1 V_2}_{^x_{LT},^{xx}_{TT}} \sin{2\varphi_2}\bigr)
    \bigr]
    \nonumber \\
    & -\frac{3}{2}\sin{2\theta_1} \sin{\varphi_1} \bigl[
          {t}^{V_1 V_2}_{^y_{LT},LL} (1-3\cos^2{\theta_2})
        -\frac{3}{2}\sin{2\theta_2} \bigl({t}^{V_1 V_2}_{^y_{LT},^x_{LT}} \cos{\varphi_2} + {t}^{V_1 V_2}_{^y_{LT},^y_{LT}} \sin{\varphi_2}\bigr)
        \nonumber \\
        &\qquad\qquad\qquad\qquad -\frac{3}{2}\sin^2{\theta_2} \bigl({t}^{V_1 V_2}_{^y_{LT},^{xx}_{TT}} \cos{2\varphi_2} + {t}^{V_1 V_2}_{^y_{LT},^{xy}_{TT}} \sin{2\varphi_2}\bigr)
    \bigr]
    \nonumber \\
    & -\frac{3}{2}\sin^2{\theta_1} \cos{2\varphi_1} \bigl[
          {t}^{V_1 V_2}_{^{xx}_{TT},LL} (1-3\cos^2{\theta_2})
        -\frac{3}{2}\sin{2\theta_2} \bigl({t}^{V_1 V_2}_{^{xx}_{TT},^x_{LT}} \cos{\varphi_2} + {t}^{V_1 V_2}_{^{xx}_{TT},^y_{LT}} \sin{\varphi_2}\bigr)
        \nonumber \\
    &\qquad\qquad\qquad\qquad -\frac{3}{2}\sin^2{\theta_2} \bigl({t}^{V_1 V_2}_{^{xx}_{TT},^{xx}_{TT}} \cos{2\varphi_2} + {t}^{V_1 V_2}_{^{xx}_{TT},^{xy}_{TT}} \sin{2\varphi_2}\bigr)
    \bigr]
    \nonumber \\
    & -\frac{3}{2}\sin^2{\theta_1} \sin{2\varphi_1} \bigl[
         {t}^{V_1 V_2}_{^{xy}_{TT},LL} (1-3\cos^2{\theta_2})
        -\frac{3}{2}\sin{2\theta_2} \bigl({t}^{V_1 V_2}_{^{xy}_{TT},^x_{LT}} \cos{\varphi_2} + {t}^{V_1 V_2}_{^{xy}_{TT},^y_{LT}} \sin{\varphi_2}\bigr)
        \nonumber \\
    &\qquad\qquad\qquad\qquad -\frac{3}{2}\sin^2{\theta_2} \bigl({t}^{V_1 V_2}_{^{xy}_{TT},^{xx}_{TT}} \cos{2\varphi_2} + {t}^{V_1 V_2}_{^{xy}_{TT},^{xy}_{TT}} \sin{2\varphi_2}\bigr)
    \bigr]
\Bigr\}.
\label{eq:A10}
\end{align}
Here, if focus only on the longitudinal components,  Eq.~(\ref{eq:A10}) is given by
\begin{align}
W_{V_1V_2}(\theta_1,\theta_2)=\frac{1}{4}\Bigl[ &
1+S_{LL}(V_1)(1-3\cos^2\theta_1)+S_{LL}(V_2)(1-3\cos^2\theta_2 )
\nonumber
\\
&+ t^{V_1 V_2}_{LL,LL} (1-3\cos^2\theta_1)(1-3\cos^2\theta_2)\Bigr],
\end{align}
where the indices $LL$, $^x_{LT}$, $^y_{LT}$, $^{xy}_{TT}$ and $^{xx}_{TT}$ denote the five independent components of the rank-2 tensor polarization, corresponding to $\hat{\Sigma}_{LL}$, $\hat{\Sigma}^x_{LT}$, $\hat{\Sigma}^y_{LT}$, $\hat{\Sigma}^{xy}_{TT}$ and $\hat{\Sigma}^{xx}_{TT}$ respectively.

\section{Hadron-hadron spin correlations in terms of spin density matrix elements}\label{appendix:hhsc}

Hadron-hadron spin correlations are defined in Eq.~(\ref{eq:hcijdef}) and can be calculated from Eq.~(\ref{eq:cijcal}). 
In this appendix, we provide the explicit expressions for the spin correlation coefficients in terms of the hadronic spin density matrix elements.
We present results for $t_{ij}^{h_1h_2} =c_{ij}^{h_1h_2}+P_{h_1}^{(i)}P_{h_2}^{(j)}$ in the following.

 \subsection{Spin-$1/2$ hyperon-spin-$1/2$ hyperon $HH$ spin correlations}
For  $J^P=(1/2)^+$ hyperon pairs $H_1 H_2$, besides those given in the text, we have
\begin{align}
t^{H_1 H_2}_{xx}
&= 2\,\Re\!\left[
 \rho^{H_1 H_2}_{--,++}
 +\rho^{H_1 H_2}_{-+,+-}
 \right],
\nonumber\\
t^{H_1 H_2}_{xy}
&= 2\,\Im\!\left[
 \rho^{H_1 H_2}_{--,++}
 -\rho^{H_1 H_2}_{-+,+-}
 \right],
\nonumber\\
t^{H_1 H_2}_{yx}
&= 2\,\Im\!\left[
 \rho^{H_1 H_2}_{--,++}
 +\rho^{H_1 H_2}_{-+,+-}
 \right],
\nonumber\\
t^{H_1 H_2}_{yy}
&= 2\,\Re\!\left[
 \rho^{H_1 H_2}_{-+,+-}
 -\rho^{H_1 H_2}_{--,++}
 \right],
\nonumber\\
t^{H_1 H_2}_{yz}
&= 2\,\Im\!\left[
 \rho^{H_1 H_2}_{-+,++}
 -\rho^{H_1 H_2}_{--,+-}
 \right],
\nonumber\\
t^{H_1 H_2}_{zy}
&= 2\,\Im\!\left[
 \rho^{H_1 H_2}_{+-,++}
 -\rho^{H_1 H_2}_{--,-+}
 \right].
\label{eq:tmunuH1H2}
\end{align}

\subsection{Spin-$1/2$ hyperon-vector meson $HV$ spin correlations}\label{appendix:correlation coefficients}

For spin-$1/2$ hyperon and spin-$1$ vector meson, there are totally $3\times8=24$ independent spin correlations
We classify them as correlations between vector and vector polarization and those between vector and tensor polarization two parts 
and present results for $t_{ij}^{h_1h_2} =c_{ij}^{h_1h_2}+P_{h_1}^{(i)}P_{h_2}^{(j)}$ in the following.

(1) The vector and vector polarization part, calculated from $\Tr [\hat\rho^{HV}\hat{\sigma}_m \otimes \hat\Sigma^n]$.

\begin{align}
        t^{HV}_{x,x}=& \sqrt{2} \,\Re \!\left[\rho^{HV}_{-0,+-1}+\rho^{HV}_{--1,+0}+\rho^{HV}_{-0,+1}+\rho^{HV}_{-1,+0} \right],
        \nonumber
        \\
        t^{HV}_{x,y}=& \sqrt{2} \, \Im \!\left[-\rho^{HV}_{-0,+-1}+\rho^{HV}_{--1,+0}+\rho^{HV}_{-0,+1}-\rho^{HV}_{-1,+0} \right],
        \nonumber
        \\
         t^{HV}_{x,z}=& 2 \,\Re \!\left[\rho^{HV}_{-1,+1}-\rho^{HV}_{--1,+-1} \right],
         \nonumber
         \\
         t^{HV}_{y,x}=& \sqrt{2} \,\Im \!\left[\rho^{HV}_{-0,+-1}+\rho^{HV}_{--1,+0}+\rho^{HV}_{-0,+1}+\rho^{HV}_{-1,+0} \right],
         \nonumber
         \\
          t^{HV}_{y,y}=& \sqrt{2} \,\Re \!\left[ \rho^{HV}_{-0,+-1}-\rho^{HV}_{--1,+0}+\rho^{HV}_{-1,+0}-\rho^{HV}_{-0,+1} \right],
          \nonumber
          \\
           t^{HV}_{y,z}=& 2 \, \Im \!\left[\rho^{HV}_{-1,+1}-\rho^{HV}_{--1,+0} \right],
           \nonumber
           \\
           t^{HV}_{z,x}=& \sqrt{2}  \, \Re \!\left[\rho^{HV}_{+0,+1}+\rho^{HV}_{+-1,+0}-\rho^{HV}_{-0,-1}-\rho^{HV}_{--1,-0} \right],
           \nonumber
           \\
            t^{HV}_{z,y}=& \sqrt{2} \, \Im \!\left[\rho^{HV}_{+0,+1}+\rho^{HV}_{+-1,+0}-\rho^{HV}_{-0,--1}-\rho^{HV}_{--1,-0} \right],
            \nonumber
            \\
            t^{HV}_{z,z}=& \rho^{HV}_{+1,+1}-\rho^{HV}_{+-1,+-1}-\rho^{HV}_{-1,-1}+\rho^{HV}_{--1,--1},
             \label{eq:B2}
    \end{align}

(2) The vector and tensor polarization part, calculated from 
$\Tr [\hat\rho^{HV}\hat{\sigma}_m \otimes \{\hat{\Sigma}_{LL}, \hat{\Sigma}^{a}_{LT}, \hat{\Sigma}^{x a}_{TT} \}]$.

\begin{align}
        t^{HV}_{x,LL}=&  \Re \!\left[\rho^{HV}_{-1,+1}-2\rho^{HV}_{-0,+0}+\rho^{HV}_{--1,+-1}\right],
        \nonumber
        \\
         t^{HV}_{x,^x_{LT}}=& \sqrt{2} \,\Re \!\left[\rho^{HV}_{-1,+0}+\rho^{HV}_{-0,+1}-\rho^{HV}_{-0,+-1}-\rho^{HV}_{--1,+0}\right],
         \nonumber
         \\
         t^{HV}_{x,^y_{LT}}=& -\sqrt{2} \, \Im \!\left[\rho^{HV}_{-1,+0}-\rho^{HV}_{-0,+1}-\rho^{HV}_{-0,+-1}+\rho^{HV}_{--1,+0}\right],
         \nonumber
         \\
         t^{HV}_{x,^{xy}_{TT}}=& 2 \,\Im
         \!\left[\rho^{HV}_{--1,+1}-\rho^{HV}_{-1,+-1}\right],
         \nonumber
         \\
         t^{HV}_{x,^{xx}_{TT}}=&  2\,\Re \!\left[\rho^{HV}_{-1,+-1}+\rho^{HV}_{--1,+1}\right],
         \nonumber
         \\
          t^{HV}_{y,LL}=&  \Im \!\left[\rho^{HV}_{-1,+1}+\rho^{HV}_{--1,+-1}-2\rho^{HV}_{-0,+0}\right],
          \nonumber
          \\
          t^{HV}_{y,^x_{LT}}=& \sqrt{2} \,\Im \!\left[\rho^{HV}_{-0,+1}+\rho^{HV}_{-1,+0}-\rho^{HV}_{--1,+0}-\rho^{HV}_{-0,+-1}\right],
          \nonumber
          \\
           t^{HV}_{y,^y_{LT}}=& \sqrt{2} \,\Re \!\left[\rho^{HV}_{-1,+0}+\rho^{HV}_{--1,+0}-\rho^{HV}_{-0,+1}-\rho^{HV}_{-0,+-1}\right],
           \nonumber
           \\
           t^{HV}_{y,^{xy}_{TT}}=&  2 \,\Re \!\left[\rho^{HV}_{-1,+-1}-\rho^{HV}_{--1,+1}\right],
           \nonumber
           \\
           t^{HV}_{y,^{xx}_{TT}}=& 2 \, \Im \! \left[\rho^{HV}_{-1,+-1}+\rho^{HV}_{--1,+1}\right],
           \nonumber
           \\
            t^{HV}_{z,LL}=& \frac{1}{2}\,\!\left[\rho^{HV}_{+1,+1}-2\rho^{HV}_{+0,+0}+\rho^{HV}_{+-1,+-1}-\rho^{HV}_{-1,-1}+2\rho^{HV}_{-0,-0}-\rho^{HV}_{--1,--1}\right],
           \nonumber
            \\
             t^{HV}_{z,^x_{LT}}=& \sqrt{2} \,\Re
             \!\left[\rho^{HV}_{+0,+1}-\rho^{HV}_{+-1,+0}-\rho^{HV}_{-0,-1}+\rho^{HV}_{--1,-0}\right],
             \nonumber
             \\
             t^{HV}_{z,^y_{LT}}=& \sqrt{2} \,\Im \!\left[\rho^{HV}_{+0,+1}+\rho^{HV}_{--1,-0}-\rho^{HV}_{+-1,+0}-\rho^{HV}_{-0,-1}\right],
             \nonumber
             \\
              t^{HV}_{z,^{xy}_{TT}}=& 2  \,\Im \!\left[\rho^{HV}_{+-1,+1}-\rho^{HV}_{--1,-1}\right],
              \nonumber
              \\
              t^{HV}_{z,^{xx}_{TT}}=& 2 \, \Re \!\left[\rho^{HV}_{+-1,+1}-\rho^{HV}_{--1,-1} \right].
              \label{eq:B3}
    \end{align}

\subsection{Vector meson-vector meson systems}
For vector meson-vector meson spin correlations, there are $56$ independent components. 
The results are given in the following.

(1) For vector and vector polarization, calculated from $\Tr [\hat\rho^{V_1V_2} \hat\Sigma^m \otimes \hat\Sigma^n]$.
\begin{align}
        t^{V_1V_2}_{x,x}=& \Re \! \left[ \rho^{V_1 V_2}_{00,11}+
        \rho^{V_1 V_2}_{01,10}+\rho^{V_1 V_2}_{0-1,10}
        +\rho^{V_1 V_2}_{00,1-1}
        +\rho^{V_1 V_2}_{-11,00}+\rho^{V_1 V_2}_{-10,01}
        +\rho^{V_1 V_2}_{-1-1,00}
        +\rho^{V_1 V_2}_{-10,0-1}
        \right],
        \nonumber
        \\
      t^{V_1V_2}_{x,y}=&  \Im \! \left[   
      \rho^{V_1 V_2}_{00,11}-\rho^{V_1 V_2}_{01,10}
      +\rho^{V_1 V_2}_{0-1,10}-\rho^{V_1 V_2}_{00,1-1}
      +\rho^{V_1 V_2}_{-10,01}-\rho^{V_1 V_2}_{-11,00}
      +\rho^{V_1 V_2}_{-1-1,00}-\rho^{V_1 V_2}_{-10,0-1}
      \right],
      \nonumber
      \\
       t^{V_1V_2}_{x,z}=& 
        \sqrt{2} \, \Re
       \! \left[   
       \rho^{V_1 V_2}_{01,11}-\rho^{V_1 V_2}_{1-1,0-1}
       +\rho^{V_1 V_2}_{-11,01}-\rho^{V_1 V_2}_{-1-1,0-1}
       \right],
       \nonumber
       \\
        t^{V_1V_2}_{y,x}=&  \Im \! \left[   
        \rho^{V_1 V_2}_{00,11}+\rho^{V_1 V_2}_{01,10}
        +\rho^{V_1 V_2}_{0-1,10}+\rho^{V_1 V_2}_{00,1-1}
        +\rho^{V_1 V_2}_{-10,01}+\rho^{V_1 V_2}_{-11,00}
        +\rho^{V_1 V_2}_{-1-1,00}+\rho^{V_1 V_2}_{-10,0-1}
        \right],
        \nonumber
        \\
         t^{V_1 V_2}_{y,y}=&  \Re \! \left[  
         -\rho^{V_1 V_2}_{00,11}+\rho^{V_1 V_2}_{01,10}
         -\rho^{V_1 V_2}_{0-1,10}+\rho^{V_1 V_2}_{00,1-1}
         -\rho^{V_1 V_2}_{-10,01}+\rho^{V_1 V_2}_{-11,00}
         -\rho^{V_1 V_2}_{-1-1,00}+\rho^{V_1 V_2}_{-10,0-1}
         \right],
         \nonumber
         \\
         t^{V_1 V_2}_{y,z}=&  \sqrt{2} \, \Im \! \left[   
         \rho^{V_1 V_2}_{01,11}-\rho^{V_1 V_2}_{0-1,1-1}
         +\rho^{V_1 V_2}_{-11,01}-\rho^{V_1 V_2}_{-1-1,0-1}
         \right],
         \nonumber
         \\
         t^{V_1 V_2}_{z,x}=& \sqrt{2} \, \Re \! \left[ 
         \rho^{V_1 V_2}_{10,11}+\rho^{V_1 V_2}_{1-1,10}
         -\rho^{V_1 V_2}_{-10,-11}-\rho^{V_1 V_2}_{-1-1,-10}
         \right],
         \nonumber
         \\
         t^{V_1 V_2}_{z,y}=& \sqrt{2} \, \Im \! \left[  
         \rho^{V_1 V_2}_{10,11}+\rho^{V_1 V_2}_{1-1,10}
         -\rho^{V_1 V_2}_{-10,-11}
         -\rho^{V_1 V_2}_{-1-1,-10}
         \right],
         \nonumber
         \\
         t^{V_1 V_2}_{z,z}=&\rho^{V_1 V_2}_{11,11}-\rho^{V_1 V_2}_{1-1,1-1}
         -\rho^{V_1 V_2}_{-11,-11}+\rho^{V_1 V_2}_{-1-1,-1-1},
         \label{eq:B4}
\end{align}

(2) For vector and tensor polarization part, calculated from $\Tr[\hat\rho^{V_1V_2} \hat\Sigma^m \otimes \{\hat{\Sigma}_{LL}, \hat{\Sigma}^{b}_{LT}, \hat{\Sigma}^{x b}_{TT} \}]$.
    \begin{align}
        t^{V_1 V_2}_{x,LL}=& \frac{\sqrt{2}}{2} \,\Re
        \! \left[  
        \rho^{V_1 V_2}_{01,11}-2\rho^{V_1 V_2}_{00,10}
        +\rho^{V_1 V_2}_{0-1,1-1}+\rho^{V_1 V_2}_{-11,01}
        -2\rho^{V_1 V_2}_{-10,00}+\rho^{V_1 V_2}_{-1-1,0-1}
        \right],
        \nonumber
        \\
        t^{V_1 V_2}_{x,^x_{LT}}=&  \Re
        \! \left[   
        \rho^{V_1 V_2}_{00,11}+\rho^{V_1 V_2}_{01,10}
        -\rho^{V_1 V_2}_{0-1,10}-\rho^{V_1 V_2}_{00,1-1}
        +\rho^{V_1 V_2}_{-10,01}+\rho^{V_1 V_2}_{-11,00}
        -\rho^{V_1 V_2}_{-1-1,00}-\rho^{V_1 V_2}_{-10,0-1}
        \right],
        \nonumber
        \\
        t^{V_1 V_2}_{x,^y_{LT}}=& \Im
        \! \left[  
        \rho^{V_1 V_2}_{00,11}-\rho^{V_1 V_2}_{01,10}
        -\rho^{V_1 V_2}_{0-1,10}+\rho^{V_1 V_2}_{00,1-1}
        +\rho^{V_1 V_2}_{-10,01}-\rho^{V_1 V_2}_{-11,00}
        -\rho^{V_1 V_2}_{-1-1,00}+\rho^{V_1 V_2}_{-10,0-1}
        \right],
        \nonumber
        \\
        t^{V_1 V_2}_{x,^{xy}_{TT}}=& \sqrt{2}\, \Im
        \! \left[   
        \rho^{V_1 V_2}_{0-1,11}-\rho^{V_1 V_2}_{01,1-1}
        +\rho^{V_1 V_2}_{-1-1,01}-\rho^{V_1 V_2}_{-11,0-1}
        \right],
        \nonumber
        \\
         t^{V_1V_2}_{x,^{xx}_{TT}}=&   \sqrt{2} \, \Re
         \! \left[  
         \rho^{V_1 V_2}_{0-1,11}+\rho^{V_1 V_2}_{01,1-1}
         +\rho^{V_1 V_2}_{-1-1,01}+\rho^{V_1 V_2}_{-11,0-1}
         \right],
         \nonumber
         \\
           t^{V_1V_2}_{y,LL}=&  \frac{\sqrt{2}}{2}\, \Im
        \! \left[ 
        \rho^{V_1 V_2}_{01,11}-2\rho^{V_1 V_2}_{00,10}
        +\rho^{V_1 V_2}_{0-1,1-1}+\rho^{V_1 V_2}_{-11,01}
        -2\rho^{V_1 V_2}_{-10,00}+\rho^{V_1 V_2}_{-1-1,0-1}
        \right],
        \nonumber
        \\
        t^{V_1 V_2}_{y,^x_{LT}}=&  \Im \! \left[  
        \rho^{V_1 V_2}_{00,11}+\rho^{V_1 V_2}_{01,10}
        -\rho^{V_1 V_2}_{0-1,10}-\rho^{V_1 V_2}_{00,1-1}
        +\rho^{V_1 V_2}_{-10,01}+\rho^{V_1 V_2}_{-11,00}
        -\rho^{V_1 V_2}_{-1-1,00}-\rho^{V_1 V_2}_{-10,0-1}
        \right],
        \nonumber
        \\
        t^{V_1V_2}_{y,^y_{LT}}=&  \Re
        \! \left[   
        -\rho^{V_1 V_2}_{00,11}+ \rho^{V_1 V_2}_{01,10}
        +\rho^{V_1 V_2}_{0-1,10}-\rho^{V_1 V_2}_{00,1-1}
        -\rho^{V_1 V_2}_{-10,01}+\rho^{V_1 V_2}_{-11,00}
        +\rho^{V_1 V_2}_{-1-1,00}-\rho^{V_1 V_2}_{-10,0-1}
        \right],
        \nonumber
        \\
        t^{V_1 V_2}_{y,^{xy}_{TT}}=& \sqrt{2} \,\Re \! \left[  
        -\rho^{V_1 V_2}_{0-1,11}+\rho^{V_1 V_2}_{01,1-1}
        -\rho^{V_1 V_2}_{-1-1,01}+\rho^{V_1 V_2}_{-11,0-1}
        \right],
        \nonumber
        \\
        t^{V_1 V_2}_{y,^{xx}_{TT}}=&\sqrt{2} \,\Im \! \left[  
        \rho^{V_1 V_2}_{0-1,11}+\rho^{V_1 V_2}_{01,1-1}
        +\rho^{V_1 V_2}_{-1-1,01}+\rho^{V_1 V_2}_{-11,0-1}
        \right],
        \nonumber
        \\
         t^{V_1 V_2}_{z,LL}=&  \frac{1}{2} \left(
        \rho^{V_1 V_2}_{11,11}-2\rho^{V_1 V_2}_{10,10}+\rho^{V_1 V_2}_{1-1,1-1}-\rho^{V_1 V_2}_{-11,-11}+2\rho^{V_1 V_2}_{-10,-10}-\rho^{V_1 V_2}_{-1-1,-1-1}  \right),
        \nonumber
        \\
         t^{V_1 V_2}_{z,^x_{LT}}=& \sqrt{2} \,\Re \! \left[ \rho^{V_1 V_2}_{10,11}
        -\rho^{V_1 V_2}_{1-1,10}-\rho^{V_1 V_2}_{-10,-11}
        +\rho^{V_1 V_2}_{-1-1,-10}
        \right],
        \nonumber
        \\
         t^{V_1V_2}_{z,^y_{LT}}=& \sqrt{2} \,\Im
         \! \left[     
         \rho^{V_1 V_2}_{10,11}-\rho^{V_1 V_2}_{1-1,10}
         -\rho^{V_1 V_2}_{-10,-11}+\rho^{V_1 V_2}_{-1-1,-10}
         \right],
         \nonumber
         \\
         t^{V_1V_2}_{z,^{xy}_{TT}}=& 2 \,\Im \! \left[    
         \rho^{V_1 V_2}_{1-1,11}-\rho^{V_1 V_2}_{-1-1,-11}
         \right],
         \nonumber
         \\
         t^{V_1 V_2}_{z,^{xx}_{TT}}=& 2 \,\Re 
         \! \left[    
         \rho^{V_1 V_2}_{1-1,11}-\rho^{V_1 V_2}_{-1-1,-11}
         \right],
         \label{eq:B5}
    \end{align}
   
(3) The tensor and vector polarization part, calculated from $\Tr [\hat\rho^{V_1V_2}\{\hat{\Sigma}_{LL}, \hat{\Sigma}^{a}_{LT}, \hat{\Sigma}^{x a}_{TT} \} \otimes \Sigma^n]$.

\begin{align}
        t^{V_1 V_2}_{LL,x}=& \frac{\sqrt{2}}{2} \,\Re
        \! \left[   
        \rho^{V_1 V_2}_{10,11}+\rho^{V_1 V_2}_{1-1,10}-2\rho^{V_1 V_2}_{00,01}-2\rho^{V_1 V_2}_{0-1,00}
        +\rho^{V_1 V_2}_{-10,-11}+\rho^{V_1 V_2}_{-1-1,-10}
        \right],
         \nonumber
        \\
        t^{V_1 V_2}_{^x_{LT},x}=& 
          \Re
        \! \left[  
        \rho^{V_1 V_2}_{00,11}+\rho^{V_1 V_2}_{01,10}+\rho^{V_1 V_2}_{0-1,10}+\rho^{V_1 V_2}_{00,1-1}
        -\rho^{V_1 V_2}_{-10,01}-\rho^{V_1 V_2}_{-11,00}-\rho^{V_1 V_2}_{-1-1,00}-\rho^{V_1 V_2}_{-10,0-1}
        \right],
         \nonumber
        \\
        t^{V_1 V_2}_{^y_{LT},x}=&   \Im \! \left[  
        \rho^{V_1 V_2}_{00,11}+\rho^{V_1 V_2}_{01,10}+
        \rho^{V_1 V_2}_{0-1,10}+\rho^{V_1 V_2}_{00,1-1}
        -\rho^{V_1 V_2}_{-10,01}-\rho^{V_1 V_2}_{-11,00}
        -\rho^{V_1 V_2}_{(-1,-1,00}-\rho^{V_1 V_2}_{-10,0-1}
        \right],
         \nonumber
        \\
        t^{V_1 V_2}_{^{xy}_{TT}, x}=&\sqrt{2} \, \Im
        \! \left[
        \rho^{V_1 V_2}_{-10,11}+\rho^{V_1 V_2}_{-11,10}
        +\rho^{V_1 V_2}_{-1-1,10}+\rho^{V_1 V_2}_{-10,1-1}
        \right],
         \nonumber
        \\
         t^{V_1 V_2}_{^{xx}_{TT},x}=&\sqrt{2} \, \Re
         \! \left[ 
         \rho^{V_1 V_2}_{-10,11}+\rho^{V_1 V_2}_{-11,10}
         +\rho^{V_1 V_2}_{-1-1,10}+\rho^{V_1 V_2}_{-10,1-1}
         \right],
         \nonumber
         \\
         t^{V_1 V_2}_{LL,y}=& \frac{\sqrt{2}}{2} \, \Im
        \! \left[ 
        \rho^{V_1 V_2}_{10,11}+\rho^{V_1 V_2}_{1-1,10}
        -2\rho^{V_1 V_2}_{00,01}-2\rho^{V_1 V_2}_{0-1,00}
        +\rho^{V_1 V_2}_{-10,-11}+\rho^{V_1 V_2}_{-1-1,-10}
        \right],
        \nonumber
        \\
        t^{V_1 V_2}_{^x_{LT},y}=& \Im
        \! \left[  
        \rho^{V_1 V_2}_{00,11}-\rho^{V_1 V_2}_{01,10}
        +\rho^{V_1 V_2}_{0-1,10}-\rho^{V_1 V_2}_{00,1-1}
        -\rho^{V_1 V_2}_{-10,01}+\rho^{V_1 V_2}_{-11,00}
        -\rho^{V_1 V_2}_{-1-1,00}+\rho^{V_1 V_2}_{-10,0-1}
        \right],
        \nonumber
        \\
        t^{V_1V_2}_{^y_{LT},y}=& \Re
        \! \left[ 
        -\rho^{V_1 V_2}_{00,11}+\rho^{V_1 V_2}_{01,10}
        -\rho^{V_1 V_2}_{0-1,10}+\rho^{V_1 V_2}_{00,1-1}
        +\rho^{V_1 V_2}_{-10,01}-\rho^{V_1 V_2}_{-11,00}
        +\rho^{V_1 V_2}_{-1-1,00}-\rho^{V_1 V_2}_{-10,0-1}
        \right],
        \nonumber
        \\
        t^{V_1 V_2}_{^{xy}_{TT},y}=&\sqrt{2}\,
        \Re
        \! \left[  
        -\rho^{V_1 V_2}_{-10,11}+\rho^{V_1 V_2}_{-11,10}
        -\rho^{V_1 V_2}_{-1-1,10}+\rho^{V_1 V_2}_{-10,1-1}
        \right],
        \nonumber
        \\
        t^{V_1 V_2}_{^{xx}_{TT},y}=& \sqrt{2} \, \Im
        \! \left[  
        \rho^{V_1 V_2}_{-10,11}-\rho^{V_1 V_2}_{-11,10}
        +\rho^{V_1 V_2}_{-1-1,10}-\rho^{V_1 V_2}_{-10,1-1}
        \right],
        \nonumber
        \\
         t^{V_1 V_2}_{LL,z}=&\frac{1}{2}  \left(  
        \rho^{V_1 V_2}_{11,11}-\rho^{V_1 V_2}_{1-1,1-1}-2\rho^{V_1 V_2}_{01,01}+2\rho^{V_1 V_2}_{0-1,0-1}
        +\rho^{V_1 V_2}_{-11,-11}-\rho^{V_1 V_2}_{-1-1,-1-1}
        \right),
        \nonumber
        \\
         t^{V_1 V_2}_{^x_{LT},z}=&  \sqrt{2} \,\Re
             \! \left[  
             \rho^{V_1 V_2}_{01,11}-\rho^{V_1 V_2}_{0-1,1-1}
             -\rho^{V_1 V_2}_{-11,01}+\rho^{V_1 V_2}_{-1-1,0-1}
             \right],
             \nonumber
        \\
        t^{V_1V_2}_{^y_{LT},z}=& \sqrt{2} \, \Im
        \! \left[  
        \rho^{V_1 V_2}_{01,11}-\rho^{V_1 V_2}_{0-1,1-1}
        -\rho^{V_1 V_2}_{-11,01}+\rho^{V_1 V_2}_{-1-1,0-1}
        \right],
        \nonumber
        \\
        t^{V_1 V_2}_{^{xy}_{TT},z}=&  2 \, \Im \! \left[  
        \rho^{V_1 V_2}_{-11,11}-\rho^{V_1 V_2}_{-1-1,1-1}
        \right],
        \nonumber
        \\
        t^{V_1 V_2}_{^{xx}_{TT},z}=&  2\, \Re \! \left[   \rho^{V_1 V_2}_{-11,11}-\rho^{V_1 V_2}_{-1-1,1-1} \right],
         \label{eq:B6}
    \end{align}
    
(4) The tensor and tensor polarization part, calculated from $\Tr [\hat\rho^{V_1V_2} \{\hat{\Sigma}_{LL}, \hat{\Sigma}^{a}_{LT}, \hat{\Sigma}^{x a}_{TT} \} \otimes \{\hat{\Sigma}_{LL}, \hat{\Sigma}^{b}_{LT}, \hat{\Sigma}^{x b}_{TT} \}]$.
\begin{align}
        t^{V_1 V_2}_{LL,LL}=& \frac{1}{4}
        \left( \rho^{V_1V_2}_{11,11}-2\rho^{V_1V_2}_{10,10}+\rho^{V_1V_2}_{1-1,1-1}-2\rho^{V_1V_2}_{01,01}
        +4\rho^{V_1V_2}_{00,00} -2\rho^{V_1V_2}_{0-1,0-1} \right.
        \nonumber
        \\
        &\left.+\rho^{V_1V_2}_{-11,-11}
        -2\rho^{V_1V_2}_{-10,-10}+\rho^{V_1V_2}_{-1-1,-1-1}
        \right),
        \nonumber
        \\
        t^{V_1 V_2}_{LL,^x_{LT}}=& \frac{\sqrt{2}}{2} \,\Re
        \! \left[  
        \rho^{V_1 V_2}_{10,11}-\rho^{V_1 V_2}_{1-1,10}-2\rho^{V_1 V_2}_{00,01}+
        2\rho^{V_1 V_2}_{0-1,00}+\rho^{V_1 V_2}_{-10,-11}-\rho^{V_1 V_2}_{-1-1,-10}
        \right],
        \nonumber
        \\
         t^{V_1 V_2}_{LL,^y_{LT}}=&  \frac{\sqrt{2}}{2} \,\Im 
        \! \left[    
        \rho^{V_1 V_2}_{10,11}-\rho^{V_1 V_2}_{1-1,10}-2\rho^{V_1 V_2}_{00,01}+2\rho^{V_1 V_2}_{0-1,00}
        +\rho^{V_1 V_2}_{-10,-11}-\rho^{V_1 V_2}_{-1-1,-10}
        \right],
        \nonumber
        \\
        t^{V_1 V_2}_{LL,^{xy}_{TT}}=&  \Im
        \! \left[   
        \rho^{V_1 V_2}_{1-1,11}-2\rho^{V_1 V_2}_{0-1,01}+
        \rho^{V_1 V_2}_{-1-1,-11}
        \right],
        \nonumber
        \\
        t^{V_1 V_2}_{LL,^{xx}_{TT}}=&\Re
        \! \left[ 
        \rho^{V_1 V_2}_{1-1,11}-2\rho^{V_1 V_2}_{0-1,01}
        +\rho^{V_1 V_2}_{-1-1,-11}
        \right],
        \nonumber
        \\
        t^{V_1 V_2}_{^x_{LT},LL}=&  \frac{\sqrt{2}}{2} \,\Re
        \! \left[   
        \rho^{V_1 V_2}_{01,11}-2\rho^{V_1 V_2}_{00,10}+\rho^{V_1 V_2}_{0-1,1-1}-\rho^{V_1 V_2}_{-11,01}
        +2\rho^{V_1 V_2}_{-10,00}-\rho^{V_1 V_2}_{-1-1,0-1}
        \right],
        \nonumber
        \\
        t^{V_1 V_2}_{^x_{LT},^x_{LT}}=&   \Re
        \! \left[ \rho^{V_1 V_2}_{00,11}+\rho^{V_1 V_2}_{01,10}
        -\rho^{V_1 V_2}_{0-1,10}-\rho^{V_1 V_2}_{00,1-1}
        -\rho^{V_1 V_2}_{-10,01}-\rho^{V_1 V_2}_{-11,00}
        +\rho^{V_1 V_2}_{-1-1,00}+\rho^{V_1 V_2}_{-10,0-1}
        \right],
        \nonumber
        \\
        t^{V_1 V_2}_{^x_{LT},^y_{LT}}=&   \Im
        \! \left[ 
        \rho^{V_1 V_2}_{00,11}-\rho^{V_1 V_2}_{01,10}
        -\rho^{V_1 V_2}_{0-1,10}+\rho^{V_1 V_2}_{00,1-1}
        -\rho^{V_1 V_2}_{-10,01}+\rho^{V_1 V_2}_{-11,00}
        +\rho^{V_1 V_2}_{-1-1,00}-\rho^{V_1 V_2}_{-10,0-1}
        \right],
        \nonumber
        \\
        t^{V_1 V_2}_{^x_{LT},^{xy}_{TT}}=&  \sqrt{2} \, \Im
        \! \left[  
        \rho^{V_1 V_2}_{0-1,11}-\rho^{V_1 V_2}_{01,1-1}
        -\rho^{V_1 V_2}_{-1-1,01}+\rho^{V_1 V_2}_{-11,0-1}
        \right],
        \nonumber
        \\
        t^{V_1 V_2}_{^x_{LT}, ^{xx}_{TT}}=& \sqrt{2} \,\Re
        \! \left[ 
        \rho^{V_1 V_2}_{0-1,11}+\rho^{V_1 V_2}_{01,1-1}
        -\rho^{V_1 V_2}_{-1-1,01}-\rho^{V_1 V_2}_{-11,0-1}
        \right],
        \nonumber
        \\
        t^{V_1 V_2}_{^y_{LT},LL}=&  \frac{\sqrt{2}}{2} \,\Im
        \! \left[   
        \rho^{V_1 V_2}_{01,11}-2\rho^{V_1 V_2}_{00,10}
        +\rho^{V_1 V_2}_{0-1,1-1}-\rho^{V_1 V_2}_{-11,01}
        +2\rho^{V_1 V_2}_{-10,00}-\rho^{V_1 V_2}_{-1-1,0-1}
        \right],
        \nonumber
        \\
        t^{V_1 V_2}_{^y_{LT},^x_{LT}}=&   \Im
        \! \left[  
        \rho^{V_1 V_2}_{00,11}+\rho^{V_1 V_2}_{01,10}
        -\rho^{V_1 V_2}_{0-1,10}-\rho^{V_1 V_2}_{00,1-1}
        -\rho^{V_1 V_2}_{-10,01}-\rho^{V_1 V_2}_{-11,00}
        +\rho^{V_1 V_2}_{-1-1,00}+\rho^{V_1 V_2}_{-10,0-1}
        \right],
        \nonumber
        \\
        t^{V_1 V_2}_{^y_{LT},^y_{LT}}=&  \Re
        \! \left[  
        -\rho^{V_1 V_2}_{00,11}+\rho^{V_1 V_2}_{01,10}
        +\rho^{V_1 V_2}_{0-1,10}-\rho^{V_1 V_2}_{00,1-1}
        +\rho^{V_1 V_2}_{-10,01}-\rho^{V_1 V_2}_{-11,00}
        -\rho^{V_1 V_2}_{-1-1,00}+\rho^{V_1 V_2}_{-10,0-1}
        \right],
        \nonumber
        \\
        t^{V_1 V_2}_{^y_{LT},^{xy}_{TT}}=& \sqrt{2} \,\Re
        \! \left[   
        -\rho^{V_1 V_2}_{0-1,11}+\rho^{V_1 V_2}_{01,1-1}
        +\rho^{V_1 V_2}_{-1-1,01}-\rho^{V_1 V_2}_{-11,0-1}
        \right],
        \nonumber
        \\
        t^{V_1 V_2}_{^{y}_{LT},^{xx}_{TT}}=& \sqrt{2}\,
        \Im
        \! \left[ 
        \rho^{V_1 V_2}_{0-1,11}+\rho^{V_1 V_2}_{01,1-1}
        -\rho^{V_1 V_2}_{-1-1,01}-\rho^{V_1 V_2}_{-11,0-1}
        \right],
        \nonumber
        \\
         t^{V_1V_2}_{^{xy}_{TT},LL}=&  \Im
        \! \left[ 
        \rho^{V_1 V_2}_{-11,11}
        -2 \rho^{V_1 V_2}_{-10,10}+\rho^{V_1 V_2}_{-1-1,1-1}
        \right],
        \nonumber
        \\
        t^{V_1 V_2}_{^{xy}_{TT},^{x}_{LT}}=& \sqrt{2} \,\Im
        \! \left[     
        \rho^{V_1 V_2}_{-10,11}
        +\rho^{V_1 V_2}_{-11,10}
        -\rho^{V_1 V_2}_{-1-1,10}
        -\rho^{V_1 V_2}_{-10,1-1}
        \right],
        \nonumber
        \\
        t^{V_1 V_2}_{^{xy}_{TT},^y_{LT}}=&  \sqrt{2} \,\Re
           \! \left[   
           -\rho^{V_1 V_2}_{-10,11}+ \rho^{V_1 V_2}_{-11,10}
           +\rho^{V_1 V_2}_{-1-1,10}-\rho^{V_1 V_2}_{-10,1-1}
           \right],
           \nonumber
           \\
           t^{V_1 V_2}_{^{xy}_{TT}, ^{xy}_{TT}}=& 2 \,\Re  \! \left[\rho^{V_1 V_2}_{-11,1-1}-\rho^{V_1 V_2}_{-1-1,11} \right],
           \nonumber
           \\
           t^{V_1 V_2}_{^{xy}_{TT}, ^{xx}_{TT}}=& 2 \,\Im 
           \! \left[  
           \rho^{V_1 V_2}_{-1-1,11}+\rho^{V_1 V_2}_{-11,1-1}
           \right],
           \nonumber
           \\
             t^{V_1 V_2}_{^{xx}_{TT},LL}=&  \Re
        \! \left[ 
        \rho^{V_1 V_2}_{-11,11}-2\rho^{V_1 V_2}_{-10,10}
        +\rho^{V_1 V_2}_{-1-1,1-1} \right],
        \nonumber
        \\
        t^{V_1 V_2}_{^{xx}_{TT},^x_{LT}}=& \sqrt{2} \,\Re
        \! \left[
        \rho^{V_1 V_2}_{-10,11}+\rho^{V_1 V_2}_{-11,10}
        -\rho^{V_1 V_2}_{-1-1,10}-\rho^{V_1 V_2}_{-10,1-1}
        \right],
        \nonumber
        \\
        t^{V_1 V_2}_{^{xx}_{TT},^y_{LT}}=& -\sqrt{2} \,\Im
        \! \left[   
        \rho^{V_1 V_2}_{-10,11}-\rho^{V_1 V_2}_{-11,10}
        -\rho^{V_1 V_2}_{-1-1,10}+\rho^{V_1 V_2}_{-10,1-1}
        \right],
        \nonumber
        \\
        t^{V_1 V_2}_{^{xx}_{TT},^{xy}_{TT}}=& 2 \,\Im
        \! \left[   
        \rho^{V_1 V_2}_{-1-1,11}-\rho^{V_1 V_2}_{-11,1-1}
        \right],
        \nonumber
        \\
        t^{V_1 V_2}_{^{xx}_{TT},^{xx}_{TT}}=& 2 \,\Re
        \! \left[   
        \rho^{V_1 V_2}_{-1-1,11}+\rho^{V_1 V_2}_{-11,1-1}
        \right].
        \label{eq:B7}
\end{align}

\section{Complete results for hadron–hadron spin correlations in the combination mechanism}\label{appendix:complete combination result}

In this appendix, we present the complete results for hadron-hadron spin correlations calculated using Eqs.(\ref{eq:rhoqneffective}) given in Sec.\ref{sec:hh}.

\subsection{Spin-$1/2$ hyperon-spin-$1/2$ hyperon $H_1H_2$ spin correlations}

For other $J^P=(1/2)^+$ hyperons $H_1 \bar{H}_2$, the spin correlations are given by

\begin{align}
    \bar t^{\Lambda \bar{H}_{112}}_{ij}=&\frac{\bar{\delta}_{\Lambda \bar{H}_{112}}}{\bar{C}_{\Lambda \bar{H}_{112}} }, \label{eq:C1}
    \\
     \bar t^{H_{112} \bar{H}_{112}}_{ij}=& \frac{\bar{\delta}_{H_{112} \bar{H}_{112}}}{\bar{C}_{H_{112} \bar{H}_{112}}},
      \label{eq:C2}
\end{align}
where the parameters $\bar{\delta}_{\Lambda \bar{H}_{112}}$, $\bar{C}_{\Lambda \bar{H}_{112}}$ and 
$\bar{\delta}_{H_{112} \bar{H}_{112}}$, $\bar{C}_{H_{112} \bar{H}_{112}}$
 are defined as follows
\begin{align}
    \bar{\delta}_{\Lambda \bar{H}_{112}}=&4 \bar{t}^{(s \bar{q}_1)}_{ij}-\bar{t}^{(s \bar{q}_2)}_{ij}+(\bar{t}^{(s \bar{q}_1 \bar{q}_1 \bar{q}_2)}_{illj}-4\bar{t}^{(s \bar{q}_1 \bar{q}_1 \bar{q}_2)}_{ijll})+(\bar{t}^{(uds \bar{q}_2)}_{llij}-4\bar{t}^{(uds \bar{q}_1)}_{llij})+4\bar{t}^{(uds \bar{q}_1 \bar{q}_1 \bar{q}_2)}_{llijkk}-\bar{t}^{(uds \bar{q}_1 \bar{q}_1 \bar{q}_2)}_{llikkj},
    \nonumber
    \\
    \bar{C}_{\Lambda \bar{H}_{112}}=&  3-3\bar{t}^{(ud)}_{ll}+\bar{t}^{(\bar{q}_1 \bar{q}_1)}_{ll}-4\bar{t}^{(\bar{q}_1 \bar{q}_2)}_{ll}+4\bar{t}^{(ud \bar{q}_1 \bar{q}_2)}_{llkk}-\bar{t}^{(ud \bar{q}_1 \bar{q}_1)}_{llkk},
  \nonumber
     \\
     \bar{\delta}_{H_{112} \bar{H}_{112}}=&16\bar{t}^{(q_1 \bar{q}_1)}_{ij}-4\bar{t}^{(q_1 \bar{q}_2)}_{ij}-4\bar{t}^{(\bar{q}_2 q_1)}_{ij}+\bar{t}^{(q_2 \bar{q}_2)}_{ij}
    +\big[ 4\bar{t}^{(q_2 \bar{q}_1 \bar{q}_1 \bar{q}_2)}_{ijll}-\bar{t}^{(q_2 \bar{q}_1 \bar{q}_1 \bar{q}_2)}_{illj}+4\bar{t}^{(q_1 \bar{q}_1 \bar{q}_1 \bar{q}_2)}_{illj}
    -16\bar{t}^{(q_1 \bar{q}_1 \bar{q}_1 \bar{q}_2)}_{ijll}
    \nonumber
    \\
    &+4\bar{t}^{(q_1 q_1 q_2 \bar{q}_1)}_{llij}-16\bar{t}^{(q_1 q_1 q_2 \bar{q}_1)}_{illj}
    +4\bar{t}^{(q_1 q_1 q_2 \bar{q}_2)}_{illj}-\bar{t}^{(q_1 q_1 q_2 \bar{q}_2)}_{llij}\big]
    +16\bar{t}^{(q_1 q_1 q_2 \bar{q}_1 \bar{q}_1 \bar{q}_2)}_{illjkk}-4\bar{t}^{(q_1 q_1 q_2 \bar{q}_1 \bar{q}_1 \bar{q}_2)}_{llijkk}
    \nonumber
    \\
    &+\bar{t}^{(q_1 q_1 q_2 \bar{q}_1 \bar{q}_1 \bar{q}_2 )}_{llikkj}-4\bar{t}^{(q_1 q_1 q_2 \bar{q}_1 \bar{q}_1 \bar{q}_2)}_{illkkj},
    \nonumber
    \\
     \bar{C}_{H_{112} \bar{H}_{112}}=&9+3(\bar{t}^{(q_1 q_1)}_{ll}+\bar{t}^{(\bar{q}_1 \bar{q}_1)}_{ll})-12(\bar{t}^{(q_1 q_2 )}_{ll}+\bar{t}^{(\bar{q}_1 \bar{q}_2)}_{ll})
   +4(4\bar{t}^{(q_1 q_2 \bar{q}_1 \bar{q}_2)}_{llkk}-\bar{t}^{(q_1 q_2 \bar{q}_1 \bar{q}_1)}_{llkk})+\bar{t}^{(q_1 q_1 \bar{q}_1 \bar{q}_1)}_{llkk}
   -4\bar{t}^{(q_1 q_1 \bar{q}_1 \bar{q}_2)}_{llkk},
    \label{eq:C3}
 \end{align}
where $i,j=x,y,z$ and the definitions of $\bar{t}^{(q_1 q_2)}_{ij}$ and $ \bar{t}^{(q_1 q_2 q_3)}_{ijk}$
are given in Eqs.(\ref{eq:tq1q2}) and (\ref{eq:tq1q2q3}) respectively, and $\bar{t} ^{(q_1 q_2 q_3 q_4)}_{ijkl}$ is given below
 \begin{align}
    \bar{t} ^{(q_1 q_2 q_3 q_4)}_{ijkl} =&\bar{c}^{(q_1 q_2 q_3 q_4)}_{ijkl}+
    \big(\bar{c}^{(q_1 q_2 q_3)}_{ijk} \bar{P}_{q_4 l}+\bar{c}^{(q_1 q_2 q_4)}_{ijl}\bar{P}_{q_3 k}
     +\bar{c}^{(q_1 q_3 q_4)}_{ikl}\bar{P}_{q_2 j}+\bar{c}^{(q_2 q_3 q_4)}_{jkl} \bar{P}_{q_1 i}\big)+\big(\bar{c}^{(q_1 q_2)}_{ij} \bar{P}_{q_3 k} \bar{P}_{q_4 l}
      \nonumber \\
     &+\bar{c}^{(q_1 q_3)}_{ik} \bar{P}_{q_2 j} \bar{P}_{q_4 l}+\bar{c}^{(q_1 q_4)}_{il} \bar{P}_{q_2 j} \bar{P}_{q_3 k}+
     \bar{c}^{(q_2 q_3)}_{jk} \bar{P}_{q_1 i} \bar{P}_{q_4 l}
    +\bar{c}^{(q_2 q_4)}_{jl} \bar{P}_{q_1 i} \bar{P}_{q_3 k}
     +\bar{c}^{(q_3 q_4)}_{kl} \bar{P}_{q_1 i} \bar{P}_{q_2 j} \big)
     \nonumber \\
     &+\bar{P}_{q_1 i} \bar{P}_{q_2 j} \bar{P}_{q_3 k} \bar{P}_{q_4 l} ,
    \label{eq:tq1q2q3q4}
\end{align}
higher-order coefficients like $\bar{t}^{(q_1 q_2 q_3 q_4 q_5)}_{ijkl m}$ and $\bar{t}^{(q_1 q_2 q_3 q_4 q_5 q_6)}_{ijkl m n}$ can be expanded in a similar manner.

\subsection{ Spin-$1/2$ hyperon-vector meson $HV$ spin correlations}

\subsubsection{The $\Lambda$-$V$ spin correlations}

For the $\Lambda$–$V$, the hyperon $\Lambda$ is formed by the combination of $u$, $d$, and $s$ quarks, 
while the vector meson $V$ originates from a quark–anti-quark pair $q_4\bar{q}_5$. 
The corresponding spin correlations are given by

(1) The vector and vector polarization part
\begin{align}
    \bar{t}^{ \Lambda V}_{m,n}
    =\frac{2}{ \bar{C}_{ \Lambda V}} 
    \left[
    \bar{t}^{(s q_4)}_{mn}+\bar{t}^{(s \bar{q}_5)}_{mn}
    -\bar{t}^{(uds q_4)}_{iimn}-\bar{t}^{(uds \bar{q}_5)}_{iimn}
    \right],
    \label{eq:C5}
\end{align}

(2) The vector and tensor polarization part

\begin{align}
        \bar{t}^{\Lambda V}_{m,LL}=&
       - \frac{1}{ \bar{C}_{\Lambda V }} \left[\bar{t}^{( s q_4 \bar{q}_5)}_{mii}-3\bar{t}^{( s q_4 \bar{q}_5)}_{mzz}-(\bar{t}^{( u d s q_4 \bar{q}_5)}_{iimjj}-3\bar{t}^{( u d s q_4 \bar{q}_5)}_{iimzz}) \right],
       \nonumber
       \\
       \bar{t}^{\Lambda V}_{m,^x_{LT}}=& \frac{2}{ \bar{C}_{\Lambda V}} 
       \left[   
       \bar{t}^{(s q_4 \bar{q}_5)}_{mxz}+
       \bar{t}^{(s q_4 \bar{q}_5)}_{mzx}
       -\bar{t}^{(uds q_4 \bar{q}_5)}_{iimxz}
       -\bar{t}^{(uds q_4 \bar{q}_5)}_{iimzx}
       \right],
       \nonumber
       \\
       \bar{t}^{\Lambda V}_{m,^y_{LT}}=& \frac{2}{ \bar{C}_{\Lambda V}} 
       \left[   
       \bar{t}^{(s q_4 \bar{q}_5)}_{myz}+
       \bar{t}^{(s q_4 \bar{q}_5)}_{mzy}
       -\bar{t}^{(uds q_4 \bar{q}_5)}_{iimyz}
       -\bar{t}^{(uds q_4 \bar{q}_5)}_{iimzy}
       \right],
       \nonumber
       \\
       \bar{t}^{\Lambda V}_{m,^{xy}_{TT}}=& \frac{2}{ \bar{C}_{\Lambda V}} 
       \left[   
       \bar{t}^{(s q_4 \bar{q}_5)}_{mxy}+
       \bar{t}^{(s q_4 \bar{q}_5)}_{myx}
       -\bar{t}^{(uds q_4 \bar{q}_5)}_{iimxy}
       -\bar{t}^{(uds q_4 \bar{q}_5)}_{iimyx}
       \right],
       \nonumber
       \\
       \bar{t}^{\Lambda V}_{m,^{xx}_{TT}}=&
       \frac{2}{\bar{C}_{\Lambda V}}
       \left[ 
       \bar{t}^{(s q_4 \bar{q}_5)}_{mxx}-\bar{t}^{(s q_4 \bar{q}_5)}_{myy}-(\bar{t}^{(uds q_4 \bar{q}_5)}_{iimxx}-\bar{t}^{(uds q_4 \bar{q}_5)}_{iimyy})
       \right],
       \label{eq:C6}
\end{align}
where the normalization constant $\bar{C}_{\Lambda V}$ is given by
\begin{align}
    \bar{C}_{\Lambda V }=3+\bar{t}^{(q_4 \bar{q}_5)}_{ii}-3\bar{t}^{(ud)}_{ii}
        -\bar{t}^{( ud q_4 \bar{q}_5)}_{iijj}.
        \label{eq:C7}
\end{align}

\subsubsection{The $H_{112}$-$V$ spin correlations}
Here $q_1+q_1+q_2\to H_{112}$, and $q_4+\bar q_5 \to V$. The corresponding spin correlation coefficients are given by

(1)  The vector and vector polarization part

\begin{align}
        \bar{t}^{H_{112} V }_{m,n}
        =& \frac{2}{ \bar{C}_{ H_{112} V}}
        \left[    
        4\bar{t}^{(q_1 q_4)}_{mn}-\bar{t}^{(q_2 q_4)}_{mn}+\bar{t}^{(q_1 q_1 q_2 q_4)}_{iimn}-4\bar{t}^{(q_1 q_1 q_2 q_4)}_{miin}+(q_4 \leftrightarrow \bar{q}_5)
        \right], 
        \label{eq:C8}
\end{align}

(2) The vector and tensor polarization part 

\begin{align}
   \bar{t}^{ H_{112} V}_{m,LL}
   =& \frac{1}{ \bar{C}_{H_{112} V}}
   \left[
   \bar{t}^{(q_2 q_4 \bar{q}_5)}_{mii}-3 \bar{t}^{(q_2 q_4 \bar{q}_5)}_{mzz}-4(\bar{t}^{(q_1 q_4 \bar{q}_5)}_{mii}-3\bar{t}^{(q_1 q_4 \bar{q}_5)}_{mzz})+4(\bar{t}^{(q_1 q_1 q_2 q_4 \bar{q}_5)}_{miijj} \right.
   \nonumber
   \\
   &\left.-3 \bar{t}^{(q_1 q_1 q_2 q_4 \bar{q}_5)}_{miizz})-(\bar{t}^{(q_1 q_1 q_2 q_4 \bar{q}_5)}_{iimjj}
   -3 \bar{t}^{(q_1 q_1 q_2 q_4 \bar{q}_5)}_{iimzz})
   \right],
   \nonumber
   \\
   \bar{t}^{ H_{112} V}_{m,^x_{LT}}
   =& \frac{2}{ \bar{C}_{H_{112} V}}
   \left[
   4\bar{t}^{(q_1 q_4 \bar{q}_5)}_{mxz}-\bar{t}^{(q_2 q_4 \bar{q}_5)}_{mxz}+
   \bar{t}^{(q_1 q_1 q_2 q_4 \bar{q}_5)}_{iimxz}
   -4 \bar{t}^{(q_1 q_1 q_2 q_4 \bar{q}_5)}_{miixz}
   +(q_4 \leftrightarrow \bar{q}_5)
   \right],
   \nonumber
   \\
   \bar{t}^{ H_{112} V}_{m,^{y}_{LT}}
   =& \frac{2}{ \bar{C}_{H_{112} V}}
   \left[
   4\bar{t}^{(q_1 q_4 \bar{q}_5)}_{myz}-\bar{t}^{(q_2 q_4 \bar{q}_5)}_{myz}+
   \bar{t}^{(q_1 q_1 q_2 q_4 \bar{q}_5)}_{iimyz}
   -4 \bar{t}^{(q_1 q_1 q_2 q_4 \bar{q}_5)}_{miiyz}
   +(q_4 \leftrightarrow \bar{q}_5)
   \right],
   \nonumber
   \\
   \bar{t}^{ H_{112} V}_{m,^{xy}_{TT}}
   =& \frac{2}{ \bar{C}_{H_{112} V}}
   \left[
   4\bar{t}^{(q_1 q_4 \bar{q}_5)}_{mxy}-\bar{t}^{(q_2 q_4 \bar{q}_5)}_{mxy}+
   \bar{t}^{(q_1 q_1 q_2 q_4 \bar{q}_5)}_{iimxy}
   -4 \bar{t}^{(q_1 q_1 q_2 q_4 \bar{q}_5)}_{miixy}
   +(q_4 \leftrightarrow \bar{q}_5)
   \right],
   \nonumber
   \\
   \bar{t}^{ H_{112} V}_{m,^{xx}_{TT}}
   =& \frac{2}{ \bar{C}_{H_{112} V}}
   \left[  
   4 \bar{t}^{(q_1 q_4 \bar{q}_5)}_{mxx}-\bar{t}^{(q_2 q_4 \bar{q}_5)}_{mxx}+\bar{t}^{(q_1 q_1 q_2 q_4 \bar{q}_5)}_{iimxx}-4\bar{t}^{(q_1 q_1 q_2 q_4 \bar{q_5})}_{miixx}-(xx \leftrightarrow yy)
   \right],
   \label{eq:C9}
\end{align}
where the normalization constant $\bar{C}_{H_{112} V}$ is given by
\begin{align}
    \bar{C}_{ H_{112} V}=9+3\bar{t}^{(q_4 \bar{q}_5)}_{ii}
    +3\bar{t}^{(q_1 q_1)}_{ii}-12 \bar{t}^{(q_1  q_2)}_{ii}
    +\bar{t}^{(q_1 q_1 q_4 \bar{q}_5)}_{iijj}
    -4 \bar{t}^{(q_1 q_2 q_4 \bar{q}_5)}_{iijj}.
    \label{eq:C10}
\end{align}

\subsection{Vector meson-vector meson $VV$ spin correlations }

In this case, we have $q_1+\bar q_2\to V_1$ and $q_3+\bar q_4\to V_2$ respectively. 
The corresponding spin correlations are given in the following.

(1) The vector and vector polarization part

\begin{align}
    \bar{t}^{V_1 V_2}_{m,n}= \frac{4}{\bar{C}_{V_1 V_2}}
    \left[ \bar{t}^{(\bar q_2 \bar q_4)}_{mn}+\bar{t}^{(\bar q_2 q_3)}_{mn}+\bar{t}^{(q_1 \bar q_4)}_{mn}+\bar{t}^{(q_1 q_3)}_{mn} \right],
    \label{eq:C11}
\end{align}

(2) The vector and tensor polarization part

\begin{align}
        \bar{t}^{V_1 V_2}_{m,LL}=&
        -\frac{2}{ \bar{C}_{V_1 V_2}}
       \left[\bar{t}^{(q_1 q_3 \bar q_4)}_{mii}+\bar{t}^{(\bar q_2 q_3 \bar q_4)}_{mii}-3(\bar{t}^{(q_1 q_3 \bar q_4)}_{mzz}+\bar{t}^{(\bar q_2 q_3 \bar q_4)}_{mzz}) \right],
       \nonumber
        \\
        \bar{t}^{V_1 V_2}_{m,^x_{LT}}=& \frac{4}{ \bar{C}_{V_1 V_2}}
        \left[   
        \bar{t}^{(\bar{q}_2 q_3 \bar{q}_4)}_{mxz}
        +\bar{t}^{(\bar{q}_2 q_3 \bar{q}_4)}_{mzx}
        +\bar{t}^{(q_1 q_3 \bar{q}_4)}_{mxz}
        +\bar{t}^{(q_1 q_3 \bar{q}_4)}_{mzx}
        \right],
        \nonumber
        \\
        \bar{t}^{V_1 V_2}_{m,^y_{LT}}=& \frac{4}{ \bar{C}_{V_1 V_2}}
        \left[   
        \bar{t}^{(\bar{q}_2 q_3 \bar{q}_4)}_{myz}
        +\bar{t}^{(\bar{q}_2 q_3 \bar{q}_4)}_{mzy}
        +\bar{t}^{(q_1 q_3 \bar{q}_4)}_{myz}
        +\bar{t}^{(q_1 q_3 \bar{q}_4)}_{mzy}
        \right],
        \nonumber
        \\
        \bar{t}^{V_1 V_2}_{m,^{xy}_{TT}}=& \frac{4}{ \bar{C}_{V_1 V_2}}
        \left[   
        \bar{t}^{(\bar{q}_2 q_3 \bar{q}_4)}_{mxy}
        +\bar{t}^{(\bar{q}_2 q_3 \bar{q}_4)}_{myx}
        +\bar{t}^{(q_1 q_3 \bar{q}_4)}_{mxy}
        +\bar{t}^{(q_1 q_3 \bar{q}_4)}_{myx}
        \right],
        \nonumber
        \\
        \bar{t}^{V_1 V_2}_{m,^{xx}_{TT}}=& \frac{4}{ \bar{C}_{V_1 V_2}}
      \left[\bar{t}^{(\bar{q}_2 q_3 \bar{q}_4)}_{mxx}-
        \bar{t}^{(\bar{q}_2 q_3 \bar{q}_4)}_{myy}
       +\bar{t}^{(q_1 q_3 \bar{q}_4)}_{mxx}
        -\bar{t}^{(q_1 q_3 \bar{q}_4)}_{myy}\right],
        \label{eq:C12}
\end{align}

(3) The tensor and vector polarization part

\begin{align}
    \bar{t}^{V_1 V_2}_{LL,n}=&-\frac{2}{ \bar{C}_{V_1 V_2}}
       \left[
     \bar{t}^{(q_1 \bar{q}_2 q_3)}_{iin}+
     \bar{t}^{(q_1 \bar{q}_2 \bar{q}_4)}_{iin}
     -3(\bar{t}^{(q_1 \bar{q}_2 q_3)}_{zzn}+\bar{t}^{(q_1 \bar{q}_2 \bar{q}_4)}_{zzn})
       \right],
       \nonumber
       \\
     \bar{t}^{V_1 V_2}_{^x_{LT},n}=&\frac{4}{ \bar{C}_{V_1 V_2}}
     \left[   
     \bar{t}^{(q_1 \bar{q}_2 q_3)}_{xzn}
     +\bar{t}^{(q_1 \bar{q}_2 \bar{q}_3)}_{xzn}
     +\bar{t}^{(q_1 \bar{q}_2 q_3)}_{zxn}
     +\bar{t}^{(q_1 \bar{q}_2 \bar{q}_4)}_{zxn}
     \right],
     \nonumber
     \\
     \bar{t}^{V_1 V_2}_{^y_{LT},n}=&\frac{4}{ \bar{C}_{V_1 V_2}}
     \left[   
     \bar{t}^{(q_1 \bar{q}_2 q_3)}_{yzn}
     +\bar{t}^{(q_1 \bar{q}_2 \bar{q}_3)}_{yzn}
     +\bar{t}^{(q_1 \bar{q}_2 q_3)}_{zyn}
     +\bar{t}^{(q_1 \bar{q}_2 \bar{q}_4)}_{zyn}
     \right],
     \nonumber
      \\
     \bar{t}^{V_1 V_2}_{^{xy}_{TT},n}=&\frac{4}{ \bar{C}_{V_1 V_2}}
     \left[   
     \bar{t}^{(q_1 \bar{q}_2 q_3)}_{xyn}
     +\bar{t}^{(q_1 \bar{q}_2 \bar{q}_3)}_{xyn}
     +\bar{t}^{(q_1 \bar{q}_2 q_3)}_{yxn}
     +\bar{t}^{(q_1 \bar{q}_2 \bar{q}_4)}_{yxn}
     \right],
     \nonumber
     \\
     \bar{t}^{V_1 V_2}_{^{xx}_{TT},n}=& \frac{4}{\bar{C}_{V_1 V_2}} 
     \left[   
     \bar{t}^{(q_1 \bar{q}_2 q_3)}_{xxn}
     -\bar{t}^{(q_1 \bar{q}_2 q_3)}_{yyn}
     +\bar{t}^{(q_1 \bar{q}_2 \bar{q}_4)}_{xxn}
     -\bar{t}^{(q_1 \bar{q}_2 \bar{q}_4)}_{xxn}
     -\bar{t}^{(q_1 \bar{q}_2 \bar{q}_4)}_{yyn}
     \right],
     \label{eq:C13}
\end{align}

(4) The tensor and tensor polarization part

\begin{align}
        \bar{t}^{V_1 V_2}_{LL,LL}=& \frac{1}{ \bar{C}_{V_1 V_2}}
        \left[\bar{t}^{(q_1 \bar q_2 q_3 \bar q_4)}_{iijj}-3\bar{t}^{(q_1 \bar q_2 q_3 \bar q_4)}_{iizz}-3(\bar{t}^{(q_1 \bar q_2 q_3 \bar q_4)}_{zzii}-3\bar{t}^{(q_1 \bar q_2 q_3 \bar q_4)}_{zzzz}) \right],
        \nonumber
        \\
        \bar{t}^{V_1 V_2}_{LL,^x_{LT}}=& -\frac{2}{ \bar{C}_{V_1 V_2}}
        \left[  
        \bar{t}^{(q_1 \bar{q}_2 q_3 \bar{q}_4)}_{iixz}+\bar{t}^{(q_1 \bar{q}_2 q_3 \bar{q}_4)}_{iizx}
        -3(\bar{t}^{(q_1 \bar{q}_2 q_3 \bar{q}_4)}_{zzxz}+\bar{t}^{(q_1 \bar{q}_2 q_3 \bar{q}_4)}_{zzzx})
        \right],
        \nonumber
        \\
        \bar{t}^{V_1 V_2}_{LL,^y_{LT}}=& -\frac{2}{ \bar{C}_{V_1 V_2}}
        \left[  
        \bar{t}^{(q_1 \bar{q}_2 q_3 \bar{q}_4)}_{iiyz}+\bar{t}^{(q_1 \bar{q}_2 q_3 \bar{q}_4)}_{iizy}
        -3(\bar{t}^{(q_1 \bar{q}_2 q_3 \bar{q}_4)}_{zzyz}+\bar{t}^{(q_1 \bar{q}_2 q_3 \bar{q}_4)}_{zzzy})
        \right],
        \nonumber
        \\
        \bar{t}^{V_1 V_2}_{LL,^{xy}_{TT}}=& -\frac{2}{ \bar{C}_{V_1 V_2}}
        \left[  
        \bar{t}^{(q_1 \bar{q}_2 q_3 \bar{q}_4)}_{iixy}+\bar{t}^{(q_1 \bar{q}_2 q_3 \bar{q}_4)}_{iiyx}
        -3(\bar{t}^{(q_1 \bar{q}_2 q_3 \bar{q}_4)}_{zzxy}+\bar{t}^{(q_1 \bar{q}_2 q_3 \bar{q}_4)}_{zzyx})
        \right],
        \nonumber
        \\
       \bar{t}^{V_1 V_2}_{LL,^{xx}_{TT}}=&-\frac{2}{ \bar{C}_{V_1 V_2}}
       \left[ 
       \bar{t}^{(q_1 \bar{q}_2 q_3 \bar{q}_4)}_{iixx}
       - \bar{t}^{(q_1 \bar{q}_2 q_3 \bar{q}_4)}_{iiyy}-3(\bar{t}^{(q_1 \bar{q}_2 q_3 \bar{q}_4)}_{zzxx}-\bar{t}^{(q_1 \bar{q}_2 q_3 \bar{q}_4)}_{zzyy})
       \right],
       \nonumber
       \\
       \bar{t}^{V_1 V_2}_{^x_{LT},LL}=&-\frac{2}{ \bar{C}_{V_1 V_2}}
       \left[   
       \bar{t}^{(q_1 \bar{q}_2 q_3 \bar{q}_4)}_{xzii}+ \bar{t}^{(q_1 \bar{q}_2 q_3 \bar{q}_4)}_{zxii}
       -3(\bar{t}^{(q_1 \bar{q}_2 q_3 \bar{q}_4)}_{xzzz}+\bar{t}^{(q_1 \bar{q}_2 q_3 \bar{q}_4)}_{zxzz})
       \right],
       \nonumber
       \\
       \bar{t}^{V_1 V_2}_{^y_{LT},LL}=&-\frac{2}{ \bar{C}_{V_1 V_2}}
       \left[   
       \bar{t}^{(q_1 \bar{q}_2 q_3 \bar{q}_4)}_{yzii}+ \bar{t}^{(q_1 \bar{q}_2 q_3 \bar{q}_4)}_{zyii}
       -3(\bar{t}^{(q_1 \bar{q}_2 q_3 \bar{q}_4)}_{yzzz}+\bar{t}^{(q_1 \bar{q}_2 q_3 \bar{q}_4)}_{zyzz})
       \right],
       \nonumber
       \\
       \bar{t}^{V_1 V_2}_{^{xy}_{TT},LL}=&-\frac{2}{ \bar{C}_{V_1 V_2}}
       \left[   
       \bar{t}^{(q_1 \bar{q}_2 q_3 \bar{q}_4)}_{xyii}+ \bar{t}^{(q_1 \bar{q}_2 q_3 \bar{q}_4)}_{yxii}
       -3(\bar{t}^{(q_1 \bar{q}_2 q_3 \bar{q}_4)}_{xyzz}+\bar{t}^{(q_1 \bar{q}_2 q_3 \bar{q}_4)}_{yxzz})
       \right],
       \nonumber
       \\
       \bar{t}^{V_1 V_2}_{^{xx}_{TT},LL}=&-\frac{2}{ \bar{C}_{V_1 V_2}}
       \left[   
        \bar{t}^{(q_1 \bar{q}_2 q_3 \bar{q}_4)}_{xxii}- \bar{t}^{(q_1 \bar{q}_2 q_3 \bar{q}_4)}_{yyii}-3( \bar{t}^{(q_1 \bar{q}_2 q_3 \bar{q}_4)}_{xxzz}- \bar{t}^{(q_1 \bar{q}_2 q_3 \bar{q}_4)}_{yyzz})
       \right],
       \nonumber
       \\
       \bar{t}^{V_1 V_2}_{^x_{LT},^x_{LT}}=&\frac{4}{ \bar{C}_{V_1 V_2}}
       \left[  
       \bar{t}^{(q_1 \bar{q}_2 q_3 \bar{q}_4)}_{xzxz}
       +\bar{t}^{(q_1 \bar{q}_2 q_3 \bar{q}_4)}_{xzzx}+\bar{t}^{(q_1 \bar{q}_2 q_3 \bar{q}_4)}_{zxxz}+\bar{t}^{(q_1 \bar{q}_2 q_3 \bar{q}_4)}_{zxzx}
       \right],
       \nonumber
       \\
       \bar{t}^{V_1 V_2}_{ ^x_{LT},^y_{LT}}=&\frac{4}{ \bar{C}_{V_1 V_2}}
       \left[  
       \bar{t}^{(q_1 \bar{q}_2 q_3 \bar{q}_4)}_{xzyz}
       +\bar{t}^{(q_1 \bar{q}_2 q_3 \bar{q}_4)}_{xzzy}+\bar{t}^{(q_1 \bar{q}_2 q_3 \bar{q}_4)}_{zxyz}+\bar{t}^{(q_1 \bar{q}_2 q_3 \bar{q}_4)}_{zxzy}
       \right],
       \nonumber
        \\
       \bar{t}^{V_1 V_2}_{^x_{LT},^{xy}_{TT}}=&\frac{4}{ \bar{C}_{V_1 V_2}}
       \left[  
       \bar{t}^{(q_1 \bar{q}_2 q_3 \bar{q}_4)}_{xzxy}
       +\bar{t}^{(q_1 \bar{q}_2 q_3 \bar{q}_4)}_{xzyx}+\bar{t}^{(q_1 \bar{q}_2 q_3 \bar{q}_4)}_{zxxy}+\bar{t}^{(q_1 \bar{q}_2 q_3 \bar{q}_4)}_{zxyx}
       \right],
       \nonumber
        \\
       \bar{t}^{V_1 V_2}_{^y_{LT},^x_{LT}}=&\frac{4}{ \bar{C}_{V_1 V_2}}
       \left[  
       \bar{t}^{(q_1 \bar{q}_2 q_3 \bar{q}_4)}_{yzxz}
       +\bar{t}^{(q_1 \bar{q}_2 q_3 \bar{q}_4)}_{yzzx}+\bar{t}^{(q_1 \bar{q}_2 q_3 \bar{q}_4)}_{zyxz}+\bar{t}^{(q_1 \bar{q}_2 q_3 \bar{q}_4)}_{zyzx}
       \right],
       \nonumber
        \\
       \bar{t}^{V_1 V_2}_{^y_{LT},^y_{LT}}=&\frac{4}{ \bar{C}_{V_1 V_2}}
       \left[  
       \bar{t}^{(q_1 \bar{q}_2 q_3 \bar{q}_4)}_{yzyz}
       +\bar{t}^{(q_1 \bar{q}_2 q_3 \bar{q}_4)}_{yzzy}+\bar{t}^{(q_1 \bar{q}_2 q_3 \bar{q}_4)}_{zyyz}+\bar{t}^{(q_1 \bar{q}_2 q_3 \bar{q}_4)}_{zyzy}
       \right],
       \nonumber
        \\
       \bar{t}^{V_1 V_2}_{^y_{LT},^{xy}_{TT}}=&\frac{4}{ \bar{C}_{V_1 V_2}}
       \left[  
       \bar{t}^{(q_1 \bar{q}_2 q_3 \bar{q}_4)}_{yzxy}
       +\bar{t}^{(q_1 \bar{q}_2 q_3 \bar{q}_4)}_{yzyx}+\bar{t}^{(q_1 \bar{q}_2 q_3 \bar{q}_4)}_{zyxy}+\bar{t}^{(q_1 \bar{q}_2 q_3 \bar{q}_4)}_{zyyx}
       \right],
       \nonumber
       \\
       \bar{t}^{V_1 V_2}_{^{xy}_{TT},^x_{LT}}=&\frac{4}{ \bar{C}_{V_1 V_2}}
       \left[  
       \bar{t}^{(q_1 \bar{q}_2 q_3 \bar{q}_4)}_{xyxz}
       +\bar{t}^{(q_1 \bar{q}_2 q_3 \bar{q}_4)}_{xyzx}+\bar{t}^{(q_1 \bar{q}_2 q_3 \bar{q}_4)}_{yxxz}+\bar{t}^{(q_1 \bar{q}_2 q_3 \bar{q}_4)}_{yxzx}
       \right],
       \nonumber
       \\
       \bar{t}^{V_1 V_2}_{^{xy}_{TT},^y_{LT}}=&\frac{4}{ \bar{C}_{V_1 V_2}}
       \left[  
       \bar{t}^{(q_1 \bar{q}_2 q_3 \bar{q}_4)}_{xyyz}
       +\bar{t}^{(q_1 \bar{q}_2 q_3 \bar{q}_4)}_{xyzy}+\bar{t}^{(q_1 \bar{q}_2 q_3 \bar{q}_4)}_{yxyz}+\bar{t}^{(q_1 \bar{q}_2 q_3 \bar{q}_4)}_{yxzy}
       \right],
       \nonumber
       \\
       \bar{t}^{V_1 V_2}_{^{xy}_{TT},^{xy}_{TT}}=&\frac{4}{ \bar{C}_{V_1 V_2}}
       \left[  
       \bar{t}^{(q_1 \bar{q}_2 q_3 \bar{q}_4)}_{xyxy}
       +\bar{t}^{(q_1 \bar{q}_2 q_3 \bar{q}_4)}_{xyyx}+\bar{t}^{(q_1 \bar{q}_2 q_3 \bar{q}_4)}_{yxxy}+\bar{t}^{(q_1 \bar{q}_2 q_3 \bar{q}_4)}_{yxyx}
       \right],
       \nonumber
       \\
       \bar{t}^{V_1 V_2}_{^x_{LT},^{xx}_{TT}}=&\frac{4}{ \bar{C}_{V_1 V_2}}
       \left[ \bar{t}^{(q_1 \bar{q}_2 q_3 \bar{q}_4)}_{xzxx}-\bar{t}^{(q_1 \bar{q}_2 q_3 \bar{q}_4)}_{xzyy}+\bar{t}^{(q_1 \bar{q}_2 q_3 \bar{q}_4)}_{zxxx}-\bar{t}^{(q_1 \bar{q}_2 q_3 \bar{q}_4)}_{zxyy}  \right],
       \nonumber
       \\
       \bar{t}^{V_1 V_2}_{^y_{LT},^{xx}_{TT}}=&\frac{4}{ \bar{C}_{V_1 V_2}}
       \left[ \bar{t}^{(q_1 \bar{q}_2 q_3 \bar{q}_4)}_{yzxx}-\bar{t}^{(q_1 \bar{q}_2 q_3 \bar{q}_4)}_{yzyy}+\bar{t}^{(q_1 \bar{q}_2 q_3 \bar{q}_4)}_{zyxx}-\bar{t}^{(q_1 \bar{q}_2 q_3 \bar{q}_4)}_{zyyy}  \right],
       \nonumber
       \\
       \bar{t}^{V_1 V_2}_{^{xy}_{TT},^{xx}_{TT}}=&\frac{4}{ \bar{C}_{V_1 V_2}}
       \left[ \bar{t}^{(q_1 \bar{q}_2 q_3 \bar{q}_4)}_{xyxx}-\bar{t}^{(q_1 \bar{q}_2 q_3 \bar{q}_4)}_{xyyy}+\bar{t}^{(q_1 \bar{q}_2 q_3 \bar{q}_4)}_{yxxx}-\bar{t}^{(q_1 \bar{q}_2 q_3 \bar{q}_4)}_{yxyy}  \right],
       \nonumber
       \\
       \bar{t}^{V_1 V_2}_{^{xx}_{TT},^x_{LT}}=&\frac{4}{ \bar{C}_{V_1 V_2}}
       \left[ 
       \bar{t}^{(q_1 \bar{q}_2 q_3 \bar{q}_4)}_{xxxz}-\bar{t}^{(q_1 \bar{q}_2 q_3 \bar{q}_4)}_{yyxz}+\bar{t}^{(q_1 \bar{q}_2 q_3 \bar{q}_4)}_{xxzx}-\bar{t}^{(q_1 \bar{q}_2 q_3 \bar{q}_4)}_{yyzx}
       \right],
       \nonumber
        \\
        \bar{t}^{V_1 V_2}_{^{xx}_{TT},^y_{LT}}=&\frac{4}{ \bar{C}_{V_1 V_2}}
       \left[ 
       \bar{t}^{(q_1 \bar{q}_2 q_3 \bar{q}_4)}_{xxyz}-\bar{t}^{(q_1 \bar{q}_2 q_3 \bar{q}_4)}_{yyyz}+\bar{t}^{(q_1 \bar{q}_2 q_3 \bar{q}_4)}_{xxzy}-\bar{t}^{(q_1 \bar{q}_2 q_3 \bar{q}_4)}_{yyzy}
       \right],
       \nonumber
        \\
        \bar{t}^{V_1 V_2}_{^{xx}_{TT},^{xy}_{TT}}=&\frac{4}{ \bar{C}_{V_1 V_2}}
       \left[ 
       \bar{t}^{(q_1 \bar{q}_2 q_3 \bar{q}_4)}_{xxxy}-\bar{t}^{(q_1 \bar{q}_2 q_3 \bar{q}_4)}_{yyxy}+\bar{t}^{(q_1 \bar{q}_2 q_3 \bar{q}_4)}_{xxyx}-\bar{t}^{(q_1 \bar{q}_2 q_3 \bar{q}_4)}_{yyyx}
       \right],
       \nonumber
        \\
        \bar{t}^{V_1 V_2}_{^{xx}_{TT},^{xx}_{TT}}=&  \frac{4}{\bar{C}_{V_1 V_2}}
        \left[\bar{t}^{(q_1 \bar{q}_2 q_3 \bar{q}_4)}_{xxxx}-\bar{t}^{(q_1 \bar{q}_2 q_3 \bar{q}_4)}_{xxyy}+\bar{t}^{(q_1 \bar{q}_2 q_3 \bar{q}_4)}_{yyyy}-\bar{t}^{(q_1 \bar{q}_2 q_3 \bar{q}_4)}_{yyxx}
        \right],
        \label{eq:C14}
\end{align}
where the  normalization constant $\bar{C}_{V_1 V_2}$ is given by
\begin{align} 
\bar{C}_{V_1 V_2}=9+3 \bar{t}^{(q_1 \bar{q}_2)}_{ii}+3\bar{t}^{(q_3 \bar{q}_4)}_{ii} +\bar{t}^{(q_1 \bar{q}_2 q_3 \bar{q}_4)}_{iijj}. 
\label{eq:C15} 
\end{align}

\end{widetext}

\end{document}